\documentclass[12pt]{iopart}

\usepackage{iopams} 
\usepackage{graphicx}
\usepackage{amssymb}   
\usepackage{mathrsfs}    
\usepackage[safe]{tipa}

\begin{document}


\title{Monochromatic knots and other unusual electromagnetic disturbances: light localised in 3D}

\author{Robert P Cameron}
\address{SUPA and Department of Physics, University of Strathclyde, Glasgow G4 0NG, UK}
\address{School of Physics and Astronomy, University of Glasgow, Glasgow G12 8QQ, UK}
\ead{robert.p.cameron@strath.ac.uk \\ www.ytilarihc.com}

\pagenumbering{gobble} 


\begin{abstract}
We introduce and examine a collection of unusual electromagnetic disturbances. Each of these is an exact, monochromatic solution of Maxwell's equations in free space with looped electric and magnetic field lines of finite extent and a localised appearance in all three spatial dimensions. Included are the first explicit examples of monochromatic electromagnetic knots. We also consider the generation of our unusual electromagnetic disturbances in the laboratory, at both low and high frequencies, and highlight possible directions for future research, including the use of unusual electromagnetic disturbances as the basis of a new form of three-dimensional display.
\end{abstract}


\newpage
\section{Introduction} 
Light is usually thought of as being localised in \textit{two} spatial dimensions but not three: one might imagine a ray of sunshine or a laser beam, for example. A pulse of light, which is necessarily polychromatic, can be fully localised in space at a given time, but spans an extended region as it propagates. 

In the present paper we show that it is, in fact, possible for monochromatic light to appear fully localised in free space at a fixed location and, furthermore, that such light can take on a variety of remarkable forms. Specifically, we introduce and examine a collection of `unusual electromagnetic disturbances', each of which is an exact, monochromatic solution of Maxwell's equations in free space \cite{Maxwell73, Cohen89, Griffiths99, Jackson99} with looped electric and magnetic field lines of finite extent and a localised appearance in all three spatial dimensions. Included are the first explicit examples of monochromatic electromagnetic knots (see below). We also consider the generation of our unusual electromagnetic disturbances in the laboratory, at both low and high frequencies, and highlight possible directions for future research, including the use of unusual electromagnetic disturbances as the basis of a new form of three-dimensional display.

Within optics the closest field of research is, perhaps, that of so-called electromagnetic knots, which has its origins in the work of Ra$\tilde{\textrm{n}}$ada \cite{Trautman77, Ranada89, Ranada90, Ranada92, Ranada95, Ranada97} and is now beginning to grow, rapidly \cite{Ivo04, Irvine08, Besieris09, Arrayas10, Irvine10, Arrayas11, Ranada12, Arrayas12, Dalhuisen12, vanEnk13, Kedia13, Swearngin14, Arrayas15, Thompson15, Hoyos15, Kholodenko16b, Kholodenko16a, Kedia16, Arrayas17a, Arrayas17b, Smith17, Alves17}. The term `electromagnetic knot' was introduced by Ra$\tilde{\textrm{n}}$ada to refer to electromagnetic disturbances for which ``the electromagnetic field as a whole is tied to itself'' \cite{Ranada90} in that ``any pair of magnetic lines or any pair of electric lines are a link'' \cite{Ranada95}. It has since been used more broadly (and accurately), however, to describe, in addition, electromagnetic disturbances for which the electric and magnetic field lines exhibit legitimately knotted topologies. We refer to \textit{any} electromagnetic disturbance for which the electric and / or magnetic field lines exhibit non-trivial topologies as an electromagnetic knot. Note that our focus here is \textit{not} upon threads of darkness \cite{Nye74, Berry01a, Leach04, Dennis10a,Dunlop17} or polarisation singularities \cite{Dunlop17, Hajnal87a, Hajnal87b, Berry01b, Bauer15}, which can also exhibit non-trivial topologies.

The present paper represents a new contribution to the field of electromagnetic knots in that some of our unusual electromagnetic disturbances might be regarded as the first explicit examples of \textit{monochromatic} electromagnetic knots: the electromagnetic knots introduced explicitly by others to date are \textit{polychromatic} \cite{Trautman77, Ranada89, Ranada90, Ranada92, Ranada95, Ranada97, Ivo04, Irvine08, Besieris09, Arrayas10, Irvine10, Arrayas11, Ranada12, Arrayas12, Dalhuisen12, vanEnk13, Kedia13, Swearngin14, Arrayas15, Thompson15, Hoyos15, Kholodenko16b, Kholodenko16a, Kedia16, Arrayas17a, Arrayas17b, Smith17, Alves17}. An obvious advantage of monochromaticity over polychromaticity is that it facilitates generation in the laboratory using lasers. It should be noted, however, that our unusual electromagnetic disturbances are qualitatively distinct from the electromagnetic knots introduced by others \cite{Trautman77, Ranada89, Ranada90, Ranada92, Ranada95, Ranada97, Ivo04, Irvine08, Besieris09, Arrayas10, Irvine10, Arrayas11, Ranada12, Arrayas12, Dalhuisen12, vanEnk13, Kedia13, Swearngin14, Arrayas15, Thompson15, Hoyos15, Kholodenko16b, Kholodenko16a, Kedia16, Arrayas17a, Arrayas17b, Smith17, Alves17}. In particular, some of our unusual electromagnetic disturbances can be thought of as (exotic) electromagnetic standing waves, which do not propagate, whereas most of the electromagnetic knots introduced by others can be regarded as electromagnetic pulses, propagating at the speed of light whilst distorting \cite{Trautman77, Ranada89, Ranada90, Ranada92, Ranada95, Ranada97, Ivo04, Irvine08, Besieris09, Arrayas10, Irvine10, Arrayas11, Ranada12, Arrayas12, Dalhuisen12, vanEnk13, Kedia13, Swearngin14, Arrayas15, Thompson15, Hoyos15, Kholodenko16b, Kholodenko16a, Kedia16, Arrayas17a, Arrayas17b, Smith17, Alves17}.

The total energy and other such properties of all monochromatic solutions to Maxwell's equations in free space diverge. Our unusual electromagnetic disturbances are no exception: for each disturbance the electric and magnetic fields exhibit a `$1/|\mathbf{r}|$' fall off as one moves away from the region of primary interest (as described in section \ref{fgh}), but this is not sufficient to render the total energy and other such properties of the disturbance finite. The total energy and other such properties can be rendered finite without dramatically altering the basic character of the disturbance, however, by introducing a distribution of frequencies with a small spread, in which case the disturbance is quasi-monochromatic rather than strictly monochromatic. We consider these subtleties in more detail in section \ref{Energyetc}.

We work in the classical domain, imagining ourselves to be in an inertial frame of reference described by right-handed Cartesian coordinates $x$, $y$ and $z$ with associated unit vectors $\hat{\pmb{x}}$, $\hat{\pmb{y}}$ and $\hat{\pmb{z}}$ and time $t$. $E_0$ is an electric field strength, $\omega_0=2\pi/T_0=c k_0= 2\pi c/\lambda_0$ is an angular frequency, $l$ and $l'$ are integers, $\phi_0$ is a geometrical angle and $\delta_0$ is a phase.


\newpage
\section{Motivation and overview} 
\label{Introduction}
In the strict absence of charge, the electric field $\mathbf{E}$ and magnetic field $\mathbf{B}$ obey Maxwell's equations in the form \cite{Maxwell73, Cohen89, Griffiths99, Jackson99}
\begin{eqnarray}
\pmb{\nabla}\cdot\mathbf{E}&=&0, \label{DivE}  \\
\pmb{\nabla}\cdot\mathbf{B}&=&0, \label{DivB} \\
\pmb{\nabla}\times\mathbf{E}&=& -\dot{\mathbf{B}} \label{Faraday-Lenz} \\
\pmb{\nabla}\times\mathbf{B}&=& \epsilon_0\mu_0\dot{\mathbf{E}}, \label{Maxwell}
\end{eqnarray}
where $\pmb{\nabla}$ is the del operator with respect to $\mathbf{r}=\hat{\pmb{x}}x+\hat{\pmb{y}}y+\hat{\pmb{z}}z$ and an overdot indicates partial differentiation with respect to $t$. Gauss's law (\ref{DivE}) together with the divergence theorem \cite{Katz79} states that the flux of $\mathbf{E}$ through any closed surface $\mathcal{S}$ at any given time is zero \cite{Griffiths99, Jackson99}:
\begin{equation}
\int_\mathcal{S}\mathbf{E}\cdot\textrm{d}^2\mathbf{r}=0.
\end{equation}
This permits \textit{two} distinct possibilities: an electric field line must extend indefinitely or else form a closed loop of finite extent. Similarly for $\mathbf{B}$.

\begin{figure}[h!]
\centering
\includegraphics[width=\linewidth]{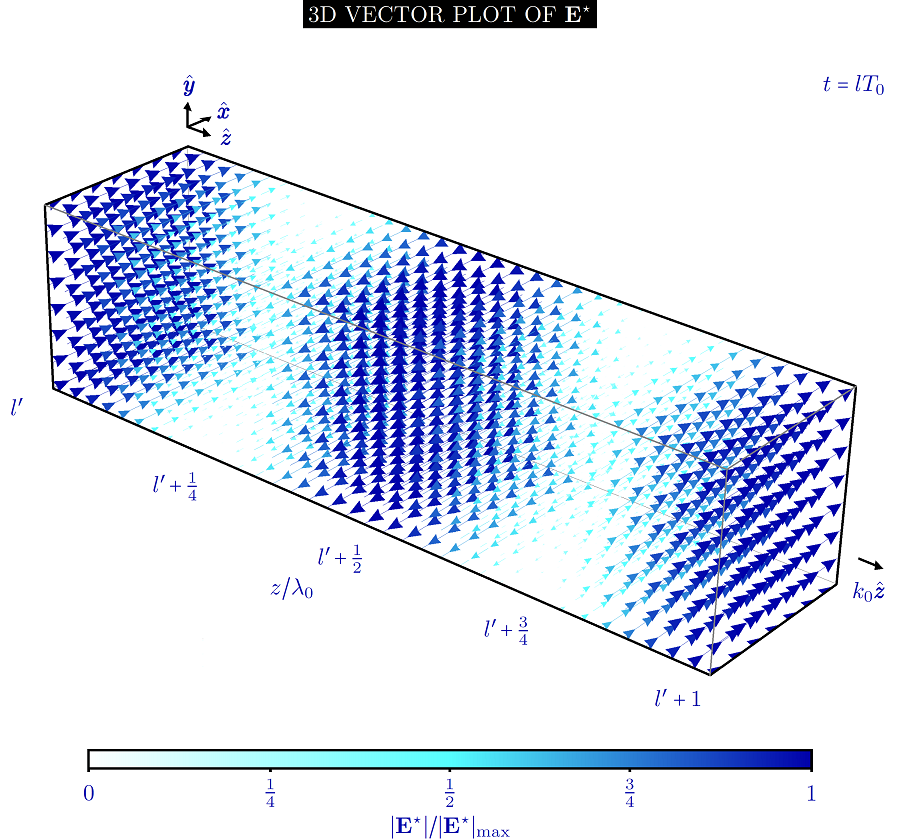}\\
\caption{\small The electric field $\mathbf{E}^{\star}$ of a linearly polarised plane electromagnetic wave (section \ref{Introduction}), depicted as a three-dimensional vector plot at an instant of time: each blue arrow represents an electric field vector, colour-coded and scaled by magnitude.} 
\label{Planewave3D}
\end{figure}

Electromagnetic disturbances with indefinitely extending electric (and magnetic) field lines are well known, a plane electromagnetic wave being a clear example. The electric field $\mathbf{E}^\star$ of the linearly polarised plane wave
\begin{eqnarray}
\mathbf{E}^{\star}&=&E_0\hat{\pmb{x}}\cos (k_0 z-\omega_0 t) \\
\mathbf{B}^{\star}&=&\frac{E_0}{c}\hat{\pmb{y}}\cos (k_0 z-\omega_0 t)
\end{eqnarray}
is depicted in Fig. \ref{Planewave3D} to illustrate this: in any given plane of constant $z$ at any given time $t\ne z/c+(1/2+l)T_0/2$ the electric field lines are straight lines of indefinite extent, parallel to the $x$ axis.

Electromagnetic disturbances with looped electric (and magnetic) field lines of finite extent are less well understood: many interesting results have been presented, in particular with regards to electromagnetic knots \cite{Trautman77, Ranada89, Ranada90, Ranada92, Ranada95, Ranada97, Ivo04, Irvine08, Besieris09, Arrayas10, Irvine10, Arrayas11, Ranada12, Arrayas12, Dalhuisen12, vanEnk13, Kedia13, Swearngin14, Arrayas15, Thompson15, Hoyos15, Kholodenko16b, Kholodenko16a, Kedia16, Arrayas17a, Arrayas17b, Smith17, Alves17}, but there remains much to be done. The author's desire to help explore this avenue led to the present paper.

The basic structure of the present paper is as follows. In section \ref{UEDs} we introduce our unusual electromagnetic disturbances and describe some of their properties. In section \ref{Energyetc} we make some general observations with regards to the energies and temporal dependencies of our unusual electromagnetic disturbances. In section \ref{Generation} we consider the generation of our unusual electromagnetic disturbances in the laboratory. In section \ref{Discuss} we highlight some possible directions for future research. 

\newpage

\newpage
The \textit{key} features common to our unusual electromagnetic disturbances are as follows.
\begin{itemize}

\item Each of our unusual electromagnetic disturbances is an \textit{exact} solution of Maxwell's equations (\ref{DivE})-(\ref{Maxwell}) and can thus exist as an entity unto itself. \\

\item Each of our unusual electromagnetic disturbances is \textit{monochromatic}. Owing to the scale invariance of (\ref{DivE})-(\ref{Maxwell}) \cite{Bessel-Hagen21}, our unusual electromagnetic disturbances are valid in any region of the electromagnetic spectrum: they can be thought of equally in the radiowave domain \cite{Maxwell65} or the gamma-ray domain \cite{Villard00a, Villard00b, Rutherford03}, for example. Our unusual electromagnetic disturbances are not electromagnetic pulses. Polychromatic variants of our unusual electromagnetic disturbances can be constructed. Save for the discussions in section \ref{Energyetc}, we refrain from pursuing these further in the present paper, however.  \\

\item Each of our unusual electromagnetic disturbances has \textit{looped} electric (and magnetic) field lines of \textit{finite} extent. Our unusual electromagnetic disturbances are not merely regions of heightened `intensity' as in a caustic \cite{Stavroudis 12a}, for example. Indeed, they cannot be understood within the framework of ray optics: interference is vital to the formation of our unusual electromagnetic disturbances, with the wavevectors, amplitudes, phases and polarisations of the plane electromagnetic waves that comprise a given disturbance all playing important roles. \\

\item Each of our unusual electromagnetic disturbances appears to be well localised in all \textit{three} spatial dimensions in that the electric (and magnetic) field strengths in the regions described are significantly larger than those found anywhere else. Our unusual electromagnetic disturbances are not propagating beams of structured light \cite{Dunlop17}, for example.

\end{itemize}

We centre our explicit discussions upon the electric field rather than the magnetic field as the electric field is often of greater importance when one comes to consider basic interactions between light and matter. Moreover, it is the looped character of the (free) electric field lines rather than magnetic field lines that is novel: $\pmb{\nabla}\cdot\mathbf{E}\ne0$ in the presence of charge whereas $\pmb{\nabla}\cdot\mathbf{B}=0$ in general, it seems \cite{Griffiths99, Jackson99, Acharya16}. As (\ref{DivE})-(\ref{Maxwell}) place $\mathbf{B}$ on equal footing with $\mathbf{E}$, the electric and magnetic properties of a disturbance can nevertheless be interchanged by performing a duality transformation: $\mathbf{E}\rightarrow c\mathbf{B}$, $\mathbf{B}\rightarrow-\mathbf{E}/c$ \cite{Jackson99, Heaviside92, Larmor97, Cameron12, Bliokh13}. 

\newpage
We use the following superscript labels to distinguish between our explicit solutions of (\ref{DivE})-(\ref{Maxwell}):
\begin{center}
\begin{tabular}{c c }
\textrm{LABEL} & \textrm{NAME} \\
$\star$ & \textrm{plane electromagnetic wave} (section \ref{Introduction}) \\
\textrm{\textipa{\textscripta}} & \textrm{electric ring} (section \ref{Rings}) \\
\textrm{\textipa{\texthtb}} & \textrm{electric globule} (section \ref{Globules}) \\
\textrm{\textipa{\texthtc}} & \textrm{electromagnetic tangle} (section \ref{Roses}) \\
\textrm{\textipa{\texthtd}} & \textrm{electric loop} (section \ref{Loops}) \\
\textrm{\textipa{\textschwa}} & \textrm{electric link} (section \ref{Links}) \\ 
\textrm{\textipa{\textdoublebaresh}} & \textrm{straight electric line} (section \ref{Lines}) \\
\textrm{\textipa{\texthtg}} & \textrm{curved electric line} (section \ref{Lines}) \\
\textrm{\textipa{\texthth}} & \textrm{electric torus knot} (section \ref{REALKnots}) \\
\textrm{\textipa{\textiota}} & \textrm{electric non-torus knot} (section \ref{REALKnots}) \\
\textrm{\textipa{\textctj}} & \textrm{electromagnetic cloud} (section \ref{Clouds}) \\
$\star\star$ & \textrm{random superposition of plane electromagnetic waves} (section \ref{Clouds})
\end{tabular}
\end{center}
and the superscript label $\textrm{LF}$ to denote the radiation generated by our electric-ring antenna (section \ref{Generation}). Magnetic versions of our electric ring, our electric globule, our electric loop, our electric link, our electric lines and our electric knots can be obtained via duality transformations. 


\newpage
\section{Unusual electromagnetic disturbances}
\label{UEDs}
In the present section we introduce our unusual electromagnetic disturbances and describe some of their properties. We distinguish between two different `kinds' of unusual electromagnetic disturbance, with the first described in section \ref{The first kind} and the second described in section \ref{The second kind}. Our unusual electromagnetic disturbances of the second kind are more elaborate than those of the first kind. 

Our approach is heuristic and we make no serious claims of completeness. Indeed, it seems clear that our unusual electromagnetic disturbances represent but a small subset of the possibilities on offer, with myriad different topologies and temporal dependencies waiting to be explored. The reader is invited to extend our approach and construct their own unusual electromagnetic disturbances, which can be shared via the online repository at www.ytilarihc.com/ued.


\subsection{The first kind}
\label{The first kind}
In the present subsection we introduce our unusual electromagnetic disturbances of the first kind. Any one of these can be regarded as a monochromatic superposition of plane electromagnetic waves propagating in every direction, each of equal amplitude, equal phase at the origin and with an equivalent polarisation, as described in more detail below. 

In section \ref{The second kind} we will employ our unusual electromagnetic disturbances of the first kind as building blocks for our unusual electromagnetic disturbances of the second kind.


\subsubsection{Construction}
\label{General recipe I} 
First, we recall that any electromagnetic disturbance can be regarded as a superposition of plane electromagnetic waves: the general solution to Maxwell's equations (\ref{DivE})-(\ref{Maxwell}) can be cast as
\begin{eqnarray}
\mathbf{E}&=&\Re\Bigg[\int_0^{2\pi}\int_0^\pi\int_0^\infty(\hat{\pmb{\varphi}}\tilde{\mathtt{B}}+\hat{\pmb{\vartheta}}\tilde{\mathtt{A}})\textrm{e}^{\textrm{i}(k\hat{\pmb{k}}\cdot\mathbf{r}-\omega t )}k^2\textrm{d}k\sin\vartheta\textrm{d}\vartheta\textrm{d}\varphi\Bigg] \label{Egeneral} \\ 
\mathbf{B}&=&\Re\Bigg[\frac{1}{c}\int_0^{2\pi}\int_0^\pi\int_0^\infty(\hat{\pmb{\varphi}}\tilde{\mathtt{A}}-\hat{\pmb{\vartheta}}\tilde{\mathtt{B}})\textrm{e}^{\textrm{i}(k\hat{\pmb{k}}\cdot\mathbf{r}-\omega t)}k^2\textrm{d}k\sin\vartheta\textrm{d}\vartheta\textrm{d}\varphi\Bigg], \label{Bgeneral}
\end{eqnarray}
where $\varphi$, $\vartheta$ and $k=\omega/c$ are spherical coordinates in reciprocal space;
\begin{eqnarray}
\hat{\pmb{\varphi}}&=&-\hat{\pmb{x}}\sin\varphi+\hat{\pmb{y}}\cos\varphi, \\
\hat{\pmb{\vartheta}}&=&\hat{\pmb{x}}\cos\varphi\cos\vartheta+\hat{\pmb{y}}\sin\varphi\cos\vartheta-\hat{\pmb{z}}\sin\vartheta \\
\hat{\pmb{k}}&=&\hat{\pmb{x}}\cos\varphi\sin\vartheta+\hat{\pmb{y}}\sin\varphi\sin\vartheta+\hat{\pmb{z}}\cos\vartheta
\end{eqnarray}
are associated unit vectors and $\tilde{\mathtt{A}}$ and $\tilde{\mathtt{B}}$ are complex functions of $\varphi$, $\vartheta$ and $k$. A particular disturbance is determined by specifying $\tilde{\mathtt{A}}$ and $\tilde{\mathtt{B}}$, which is equivalent to specifying the so-called normal variables: `$\tilde{\pmb{\alpha}}=-2\textrm{i}\sqrt{\pi^3\epsilon_0} (\hat{\pmb{\varphi}}\tilde{\mathtt{B}}+\hat{\pmb{\vartheta}}\tilde{\mathtt{A}})\textrm{e}^{-\textrm{i}\omega t}/\sqrt{\hbar\omega}$' according to the formalism described in \cite{Cohen89}, for example. An alternative expansion of the electromagnetic field, in terms of multipolar rather than plane waves, is outlined in
\ref{Multipolar}.

Next, we specialise to monochromatic disturbances by taking $\tilde{\mathtt{A}}=E_0\tilde{\mathtt{A}}' \delta(k-k_0)/2\pi k_0^2$ and $\tilde{\mathtt{B}}=E_0\tilde{\mathtt{B}}'\delta(k-k_0)/2\pi k_0^2$ without further loss of generality, where $\tilde{\mathtt{A}}'$ and $\tilde{\mathtt{B}}'$ are complex functions of $\varphi$ and $\vartheta$ only and $\delta(K)$ is the Dirac delta function: each plane wave has the same angular frequency $\omega_0=ck_0$. (\ref{Egeneral}) and (\ref{Bgeneral}) thus reduce to
\begin{eqnarray}
\mathbf{E}&=&\Re\Bigg[\frac{E_0}{2\pi}\int_0^{2\pi}\int_0^\pi(\hat{\pmb{\varphi}}\tilde{\mathtt{B}}'+\hat{\pmb{\vartheta}}\tilde{\mathtt{A}}')\textrm{e}^{\textrm{i}k_0\hat{\pmb{k}}\cdot\mathbf{r}}\sin\vartheta\textrm{d}\vartheta \textrm{d}\varphi\textrm{e}^{-\textrm{i}\omega_0 t}\Bigg] \label{Emono} \\
\mathbf{B}&=&\Re\Bigg[\frac{E_0}{2\pi c} \int_0^{2\pi}\int_0^\pi (\hat{\pmb{\varphi}}\tilde{\mathtt{A}}'-\hat{\pmb{\vartheta}}\tilde{\mathtt{B}}') \textrm{e}^{\textrm{i}k_0\hat{\pmb{k}}\cdot\mathbf{r}}\sin\vartheta\textrm{d}\vartheta \textrm{d}\varphi\textrm{e}^{-\textrm{i}\omega_0 t}\Bigg]. \label{Bmono}
\end{eqnarray}
A particular monochromatic disturbance is determined by specifying $\tilde{\mathtt{A}}'$ and $\tilde{\mathtt{B}}'$.

Finally, we obtain our unusual electromagnetic disturbances of the first kind by taking $\tilde{\mathtt{A}}'=\tilde{\mathtt{A}}'_0$ and $\tilde{\mathtt{B}}'=\tilde{\mathtt{B}}'_0$, where $\tilde{\mathtt{A}}'_0$ and $\tilde{\mathtt{B}}'_0$ are complex constants, independent of $\varphi$ and $\vartheta$: plane waves propagating in every direction, each of equal amplitude, equal phase at the origin and with an equivalent polarisation relative to $\hat{\pmb{\varphi}}$ and $\hat{\pmb{\vartheta}}$. (\ref{Emono}) and (\ref{Bmono}) thus reduce to
\begin{eqnarray}
\mathbf{E}&=&\Re\Bigg\{E_0\left[\textrm{i}\tilde{\mathtt{B}}'_0\hat{\pmb{\phi}}f+\tilde{\mathtt{A}}'_0\left(\hat{\pmb{s}}g+\hat{\pmb{z}}h\right)\right]\textrm{e}^{-\textrm{i}\omega_0 t}\Bigg\} \label{UEDe1} \\
\mathbf{B}&=&\Re \Bigg\{\frac{E_0}{c}\left[\textrm{i}\tilde{\mathtt{A}}'_0 \hat{\pmb{\phi}}f-\tilde{\mathtt{B}}'_0\left(\hat{\pmb{s}}g+\hat{\pmb{z}}h\right)\right]\textrm{e}^{-\textrm{i}\omega_0 t}\Bigg\} \label{UEDb1}
\end{eqnarray}
with the modulating scalar fields
\begin{eqnarray}
f&=&\int_0^\pi \textrm{J}_1 (k_0\sin\vartheta s)\cos \left(k_0\cos\vartheta z\right)\sin\vartheta\textrm{d}\vartheta, \label{fintegral} \\
g&=&-\int_0^\pi \textrm{J}_1 (k_0\sin\vartheta s)\sin \left(k_0\cos\vartheta z \right)  \sin\vartheta \cos\vartheta \textrm{d}\vartheta \label{gintegral} \\
h&=&-\int_0^\pi \textrm{J}_0 (k_0\sin\vartheta s)\cos\left(k_0\cos\vartheta z \right)\sin^2\vartheta \textrm{d}\vartheta, \label{hintegral}
\end{eqnarray}
where $\textrm{J}_\alpha(X)$ is the Bessel function of the first kind of order $\alpha$ and $\phi$, $s$ and $z$ are cylindrical coordinates defined such that
\begin{eqnarray}
x&=&s\cos\phi \\
y&=& s\sin\phi
\end{eqnarray}
with associated unit vectors 
\begin{eqnarray}
\hat{\pmb{\phi}}&=&-\hat{\pmb{x}}\sin\phi+\hat{\pmb{y}}\cos\phi \\
\hat{\pmb{s}}&=&\hat{\pmb{x}}\cos\phi+\hat{\pmb{y}}\sin\phi.
\end{eqnarray}
A particular unusual electromagnetic disturbance of the first kind is determined by specifying $\tilde{\mathtt{A}}'_0$ and $\tilde{\mathtt{B}}'_0$. We give explicit examples in section \ref{Rings}, section \ref{Globules} and section \ref{Roses}.


\subsubsection{The modulating scalar fields $f$, $g$ and $h$}
\label{fgh}
The forms of our unusual electromagnetic disturbances of the first kind (and, by extension, our unusual electromagnetic disturbances of the second kind (section \ref{The second kind})) derive from those of the modulating scalar fields $f$, $g$ and $h$ as seen in (\ref{UEDe1}) and (\ref{UEDb1}). We therefore examine $f$, $g$ and $h$ here for reference in what follows.

\begin{figure}[h!]
\centering
\includegraphics[width=\linewidth]{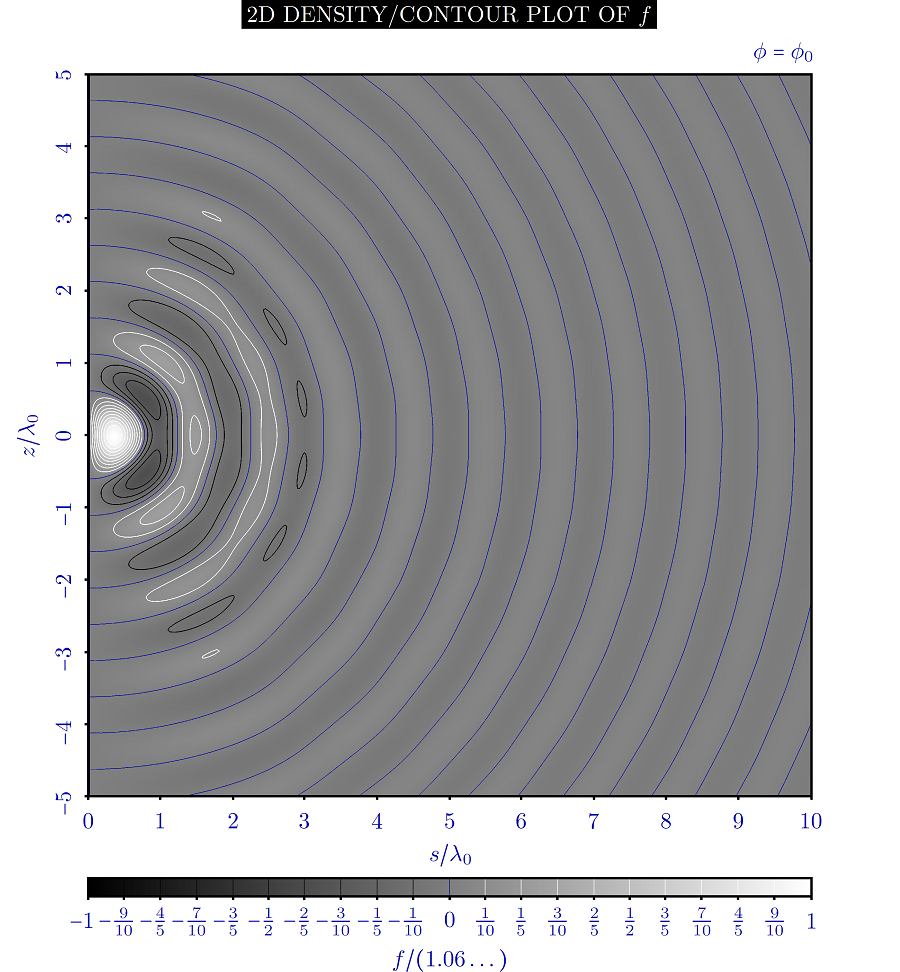}\\
\caption{\small A two-dimensional density plot of the modulating scalar field $f$ in greyscale, with contours for which $f>0$ overlaid in white, contours for which $f=0$ overlaid in blue and contours for which $f<0$ overlaid in black.} 
\label{fDECAY}
\end{figure}

\begin{figure}[h!]
\centering
\includegraphics[width=\linewidth]{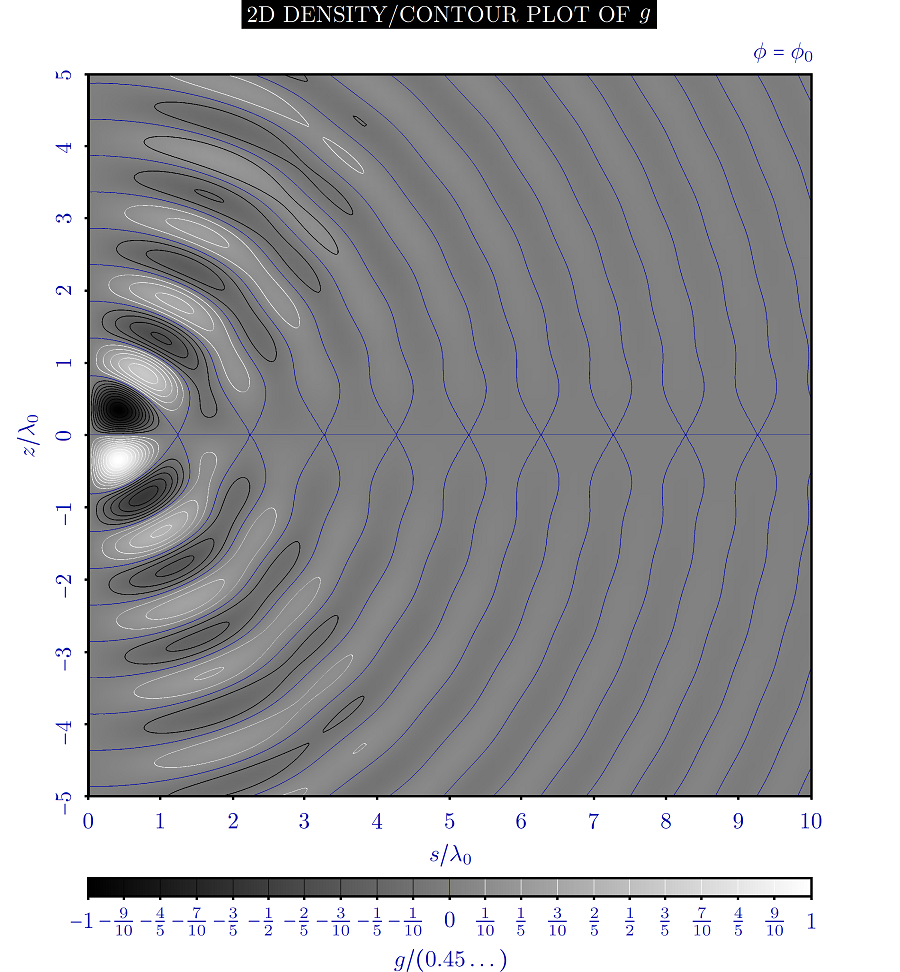}\\
\caption{\small A two-dimensional density plot of the modulating scalar field $g$ in greyscale, with contours for which $g>0$ overlaid in white, contours for which $g=0$ overlaid in blue and contours for which $g<0$ overlaid in black.} 
\label{gDECAY}
\end{figure}

\begin{figure}[h!]
\centering
\includegraphics[width=\linewidth]{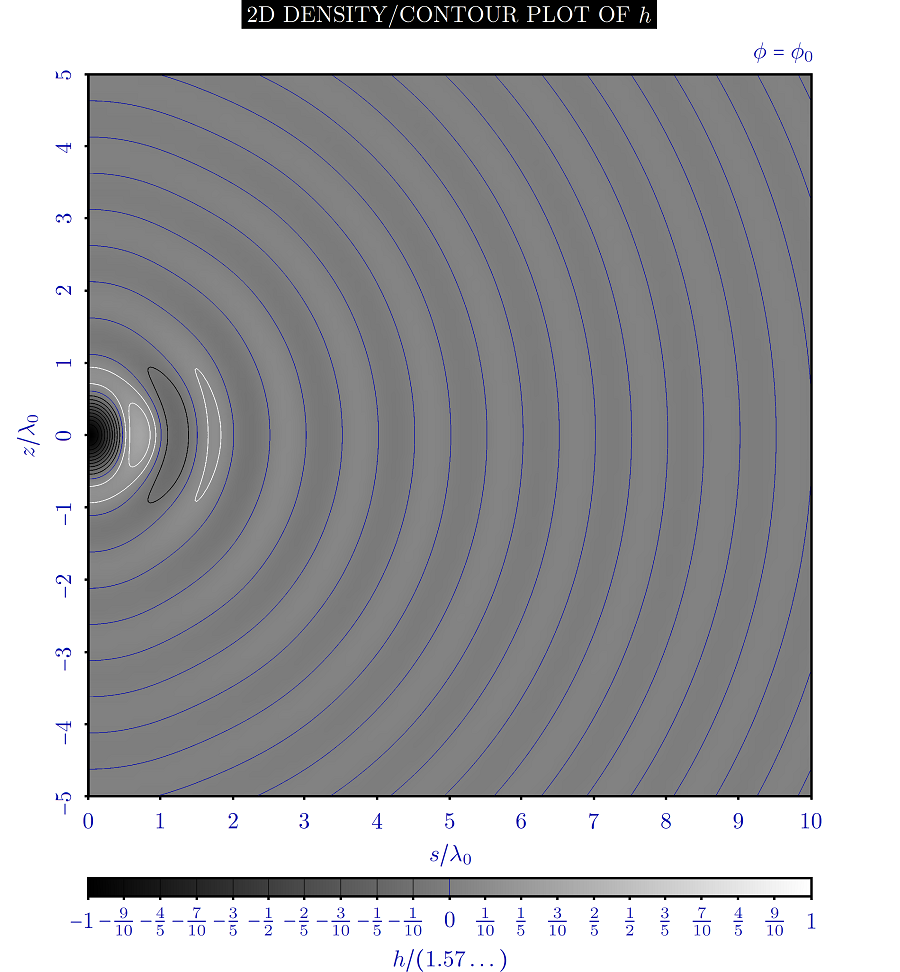}\\
\caption{\small A two-dimensional density plot of the modulating scalar field $h$ in greyscale, with contours for which $h>0$ overlaid in white, contours for which $h=0$ overlaid in blue and contours for which $h<0$ overlaid in black.} 
\label{hDECAY}
\end{figure}

$f$, $g$ and $h$ are cylindrically symmetric: they do not depend on $\phi$. Furthermore, it can be seen by inspecting (\ref{fintegral}), (\ref{gintegral}) and (\ref{hintegral}) that $f$ is even in $z$: $f(z)=f(-z)$, $g$ is odd in $z$: $g(z)=-g(-z)$ and $h$ is even in $z$: $h(z)=h(-z)$. More generally, the variation of $f$, $g$ and $h$ with $s$ and $z$ can be explored by numerical integration. The results obtained near the origin for $f$ are depicted in Fig. \ref{fDECAY}, those for $g$ are depicted in Fig. \ref{gDECAY} and those for $h$ are depicted in Fig. \ref{hDECAY}. Note the following features.
\begin{itemize}
\item The $z$ axis is a nodal line of $f$: $f(s=0,z)=0$. The highest peak of $|f|$ can be found at $s=(0.34\dots)\lambda_0$ and $z=0$, where $f=1.06\dots$. The second highest peaks of $|f|$ can be found at $s=(0.78\dots)\lambda_0$ and $z=\pm(0.52\dots)\lambda_0$, where $f=-0.41\dots$. The third highest peak of $|f|$ can be found at $s=(0.97\dots)\lambda_0$ and $z=0$, where $f=0.28\dots$. \\
\item The $z$ axis is a nodal line of $g$ and the $z=0$ plane is a nodal plane: $g(s=0,z)=g(s,z=0)=0$. The highest peaks of $|g|$ can be found at $s=(0.41\dots)\lambda_0$ and $z=\pm(0.34\dots)\lambda_0$, where $g=\mp0.44\dots$. The second highest peaks of $|g|$ can be found at $s=(0.75\dots)\lambda_0$ and $z=\pm(0.85\dots)\lambda_0$, where $g=\pm0.25\dots$. The third highest peaks of $|g|$ can be found at $s=(1.00\dots)\lambda_0$ and $z=\pm(1.34\dots)\lambda_0$, where $g=\mp0.18\dots$.  \\
\item The highest peak of $|h|$ can be found at $s=0$ and $z=0$, where $h=-1.57\dots$. The second highest peak of $|h|$ can be found at $s=(0.69\dots)\lambda_0$ and $z=0$, where $h=0.47\dots$. The third highest peak of $|h|$ can be found at $s=(1.22\dots)\lambda_0$ and $z=0$, where $h=-0.26\dots$.  \\
\end{itemize}
Further numerical investigation reveals that $f$, $g$ and $h$ each tend towards a form that is at least qualitatively similar to a sinusoidal undulation modulated by a $1/|\mathbf{r}|$ fall off as $|\mathbf{r}|\rightarrow\infty$ in any non-trivial direction. This is more dramatic than the analogous $1/\sqrt{X}$ fall off inherent to each of the $\textrm{J}_\alpha(X)$ seen in the integrands of (\ref{fintegral}), (\ref{gintegral}) and (\ref{hintegral}).

$f$, $g$ and $h$ are related by
\begin{eqnarray}
-k_0 f &=& \frac{\partial g}{\partial z}-\frac{\partial h}{\partial s}, \\
-k_0 g&=&-\frac{\partial f}{\partial z} \\
-k_0 h&=& \frac{1}{s}f +\frac{\partial f}{\partial s},
\end{eqnarray}
in accord with the Faraday-Lenz law (\ref{Faraday-Lenz}) and the Amp\`{e}re-Maxwell law (\ref{Maxwell}).

\newpage

\newpage

\newpage


\newpage
\subsubsection{Electric ring}
\label{Rings}
\begin{figure}[h!]
\centering
\includegraphics[width=\linewidth]{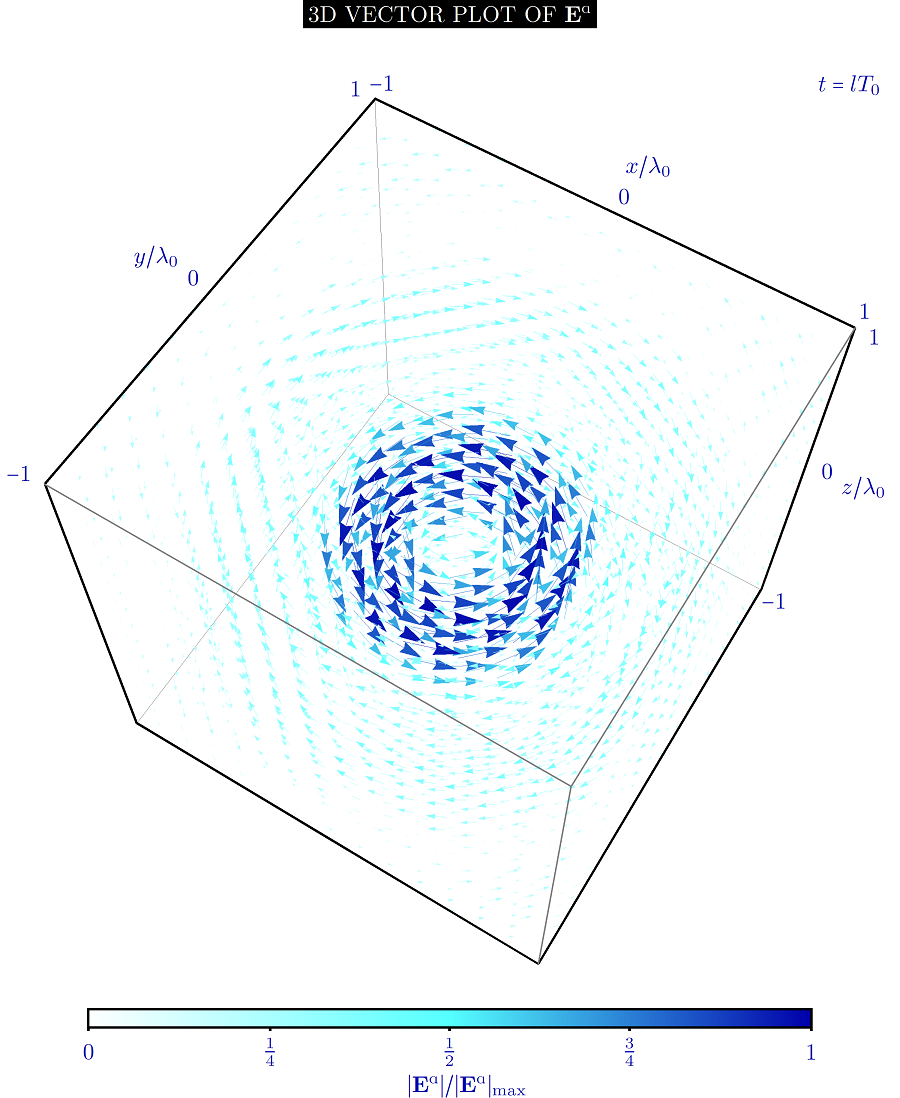}\\
\caption{\small The electric field $\mathbf{E}^\textrm{\textipa{\textscripta}}$ of our electric ring (section \ref{Rings}), depicted as a three-dimensional vector plot at an instant of time: each blue arrow represents an electric field vector, colour-coded and scaled by magnitude.} 
\label{Ring3D}
\end{figure}

\begin{figure}[h!]
\centering
\includegraphics[width=\linewidth]{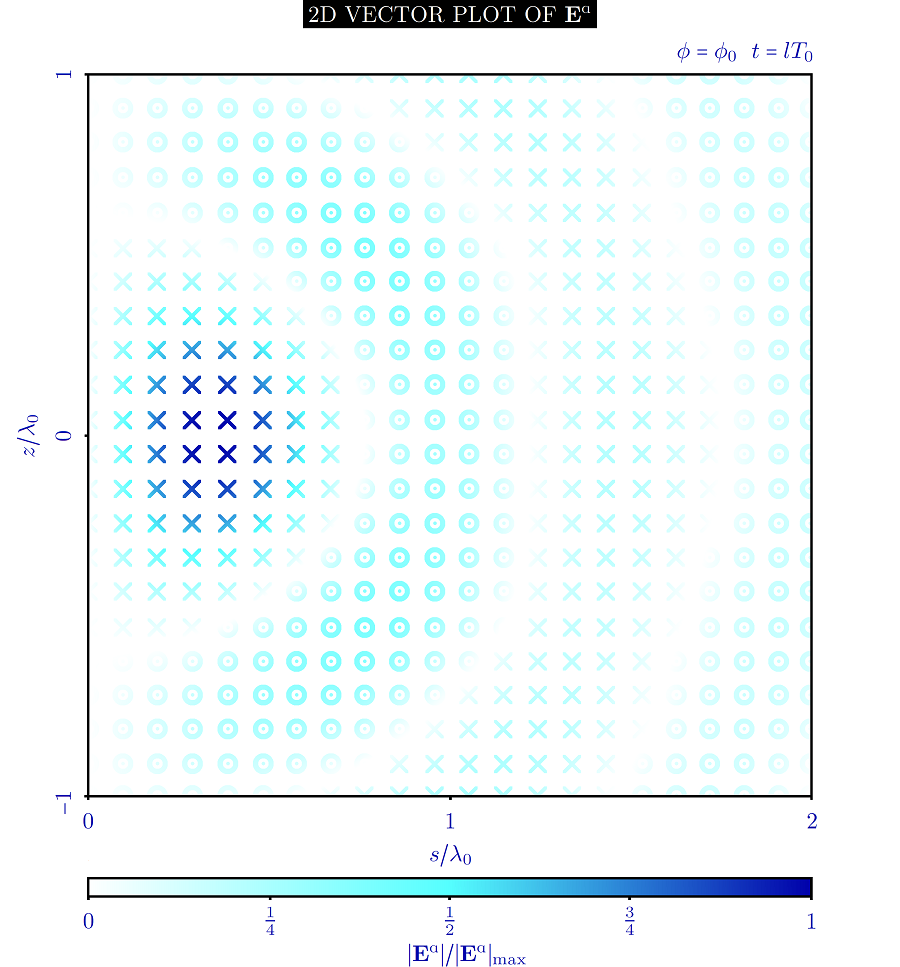}\\
\caption{\small The electric field $\mathbf{E}^\textrm{\textipa{\textscripta}}$ of our electric ring (section \ref{Rings}), depicted as a two-dimensional vector plot at an instant of time: each blue cross ($\mathbf{E}^\textrm{\textipa{\textscripta}}\cdot\hat{\pmb{\phi}}>0$) or circle ($\mathbf{E}^\textrm{\textipa{\textscripta}}\cdot\hat{\pmb{\phi}}<0$) represents an electric field vector, colour-coded by magnitude.} 
\label{Ring2D}
\end{figure}

Taking $\tilde{\mathtt{A}}'_0=0$ and $\tilde{\mathtt{B}}'_0=-\textrm{i}$ in (\ref{UEDe1}) and (\ref{UEDb1}) corresponds to having each plane wave linearly polarised parallel to $\hat{\pmb{\varphi}}$ and gives our `electric ring':
\begin{eqnarray}
\mathbf{E}^\textrm{\textipa{\textscripta}}&=&E_0\hat{\pmb{\phi}} f \cos \left(\omega_0 t\right) \\
\mathbf{B}^\textrm{\textipa{\textscripta}}&=&\frac{E_0}{c}\left(\hat{\pmb{s}}g+\hat{\pmb{z}}h\right)\sin\left(\omega_0 t\right),
\end{eqnarray}
so-named because when $t\ne(1/2+l)T_0/2$ some of the electric field lines form a particularly prominent ring-like feature near the origin. The electric field $\mathbf{E}^\textrm{\textipa{\textscripta}}$ of our electric ring is depicted in Fig. \ref{Ring3D} and Fig. \ref{Ring2D}. Note that the magnetic field $\mathbf{B}^\textrm{\textipa{\textscripta}}$ of our electric ring is equivalent to the electric field $\mathbf{E}^\textrm{\texthtb}$ of our electric globule (section \ref{Globules}) in that $\mathbf{B}^\textrm{\textipa{\textscripta}}=\mathbf{E}^\textrm{\textipa{\texthtb}}/c$.

Our electric ring is qualitatively similar to the electromagnetic disturbance described in \cite{Chubykalo02} in that both are monochromatic and exhibit azimuthally directed electric field vectors. The two differ, however, in their spatial dependence and one might argue that the particularly prominent ring-like feature of our electric ring is more pronounced than the analogous feature in \cite{Chubykalo02}. See also \cite{Arnhoff92}.


\subsubsection{Electric globule}
\label{Globules}
\begin{figure}[h!]
\centering
\includegraphics[width=\linewidth]{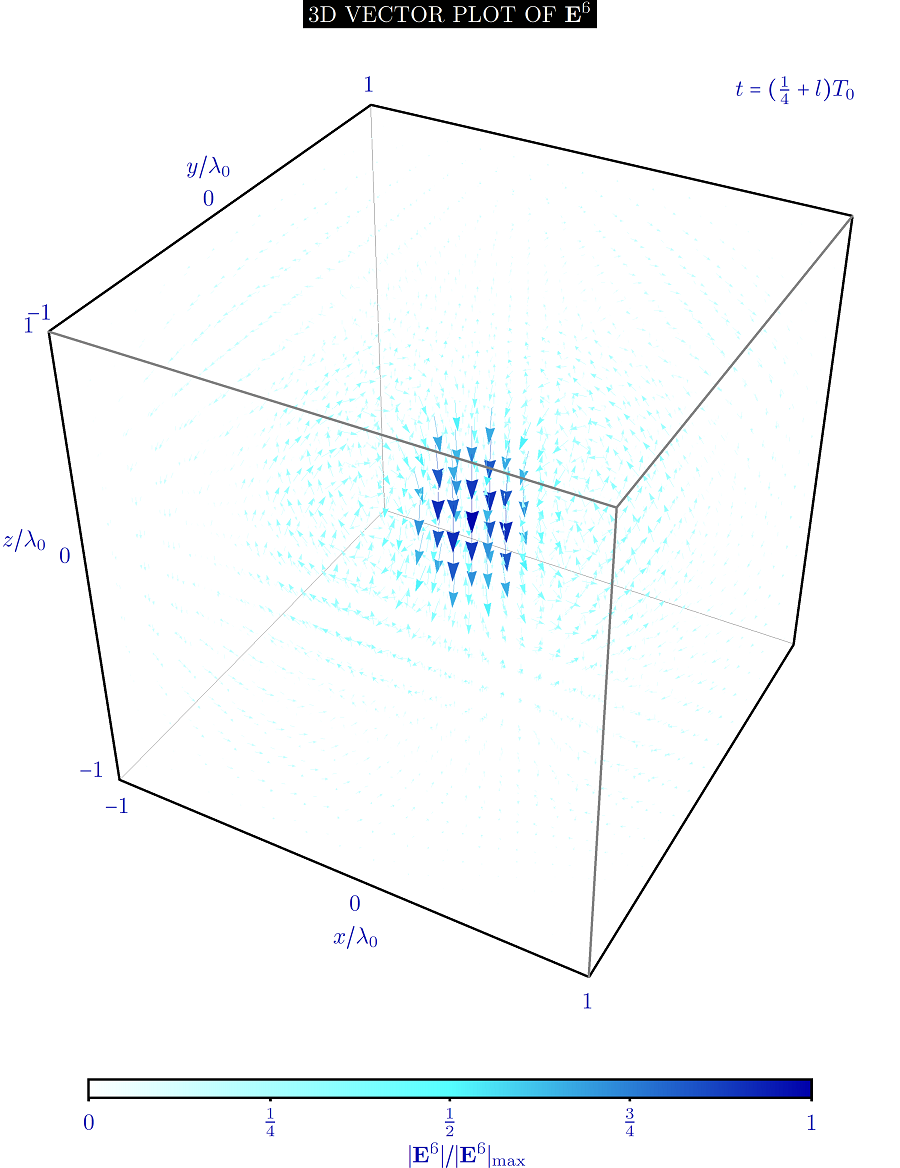}\\
\caption{\small The electric field $\mathbf{E}^\textrm{\textipa{\texthtb}}$ of our electric globule (section \ref{Globules}), depicted as a three-dimensional vector plot at an instant of time: each blue arrow represents an electric field vector, colour-coded and scaled by magnitude.} 
\label{Globule3D}
\end{figure}

\begin{figure}[h!]
\centering
\includegraphics[width=\linewidth]{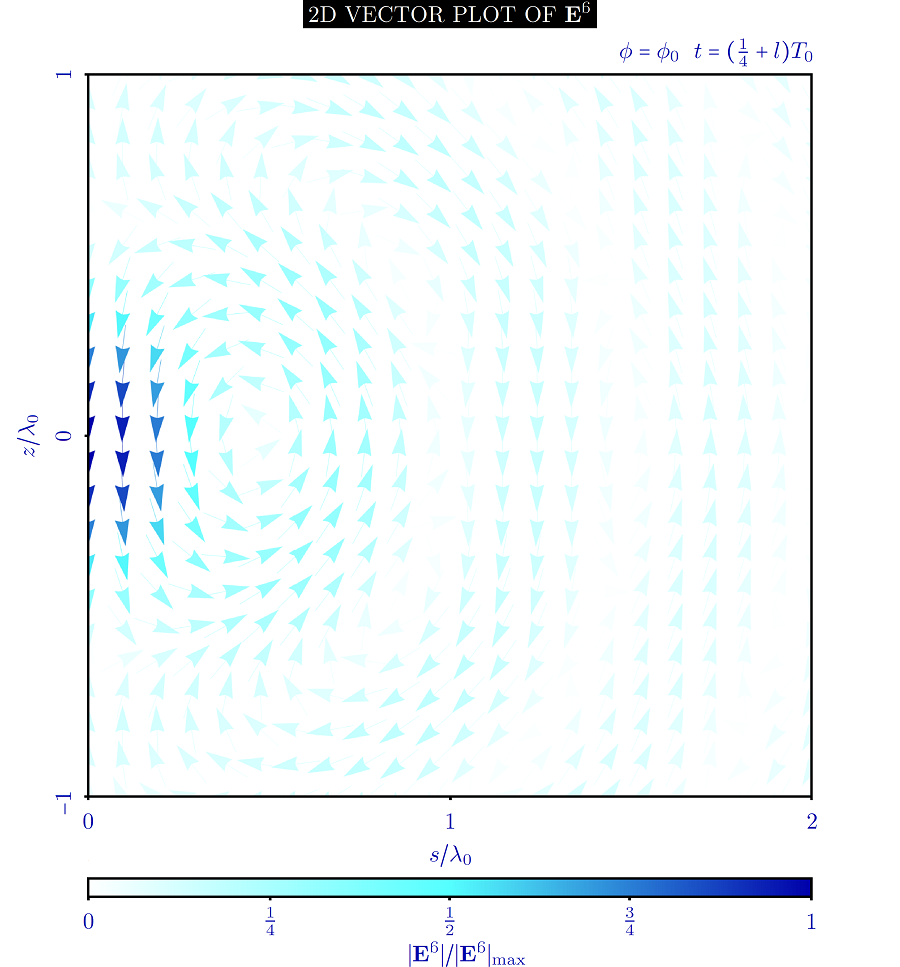}\\
\caption{\small The electric field $\mathbf{E}^\textrm{\textipa{\texthtb}}$ of our electric globule (section \ref{Globules}), depicted as a two-dimensional vector plot at an instant of time: each blue arrow represents an electric field vector, colour-coded by magnitude.} 
\label{Globule2D}
\end{figure}

Taking $\tilde{\mathtt{A}}'_0=\textrm{i}$ and $\tilde{\mathtt{B}}'_0=0$ in (\ref{UEDe1}) and (\ref{UEDb1}) corresponds to having each plane wave linearly polarised parallel to $\hat{\pmb{\vartheta}}$ and gives our `electric globule':
\begin{eqnarray}
\mathbf{E}^\textrm{\textipa{\texthtb}}&=&E_0\left(\hat{\pmb{s}}g+\hat{\pmb{z}}h\right)\sin\left(\omega_0 t\right) \\
\mathbf{B}^\textrm{\textipa{\texthtb}}&=&-\frac{E_0}{c}\hat{\pmb{\phi}}f\cos\left(\omega_0 t\right),
\end{eqnarray}
so-named because when $t\ne l T_0/2$ some of the electric field lines wind polloidally and cluster near the origin, giving the disturbance a globular appearance. The electric field $\mathbf{E}^\textrm{\textipa{\texthtb}}$ of our electric globule is depicted in Fig. \ref{Globule3D} and Fig. \ref{Globule2D}. Note that the magnetic field $\mathbf{B}^\textrm{\textipa{\texthtb}}$ of our electric globule is equivalent to the electric field $\mathbf{E}^\textrm{\textipa{\textscripta}}$ of our electric ring (section \ref{Rings}) in that $\mathbf{B}^\textrm{\textipa{\texthtb}}=-\mathbf{E}^\textrm{\textipa{\textscripta}}/c$.

Our electric globule is qualitatively similar to a duality-transformed version of the electromagnetic disturbance described in \cite{Chubykalo02}. See also \cite{Arnhoff92}.

\newpage

\newpage

\newpage

\newpage


\newpage
\subsubsection{Electromagnetic tangle}
\label{Roses}
\begin{figure}[h!]
\centering
\includegraphics[width=\linewidth]{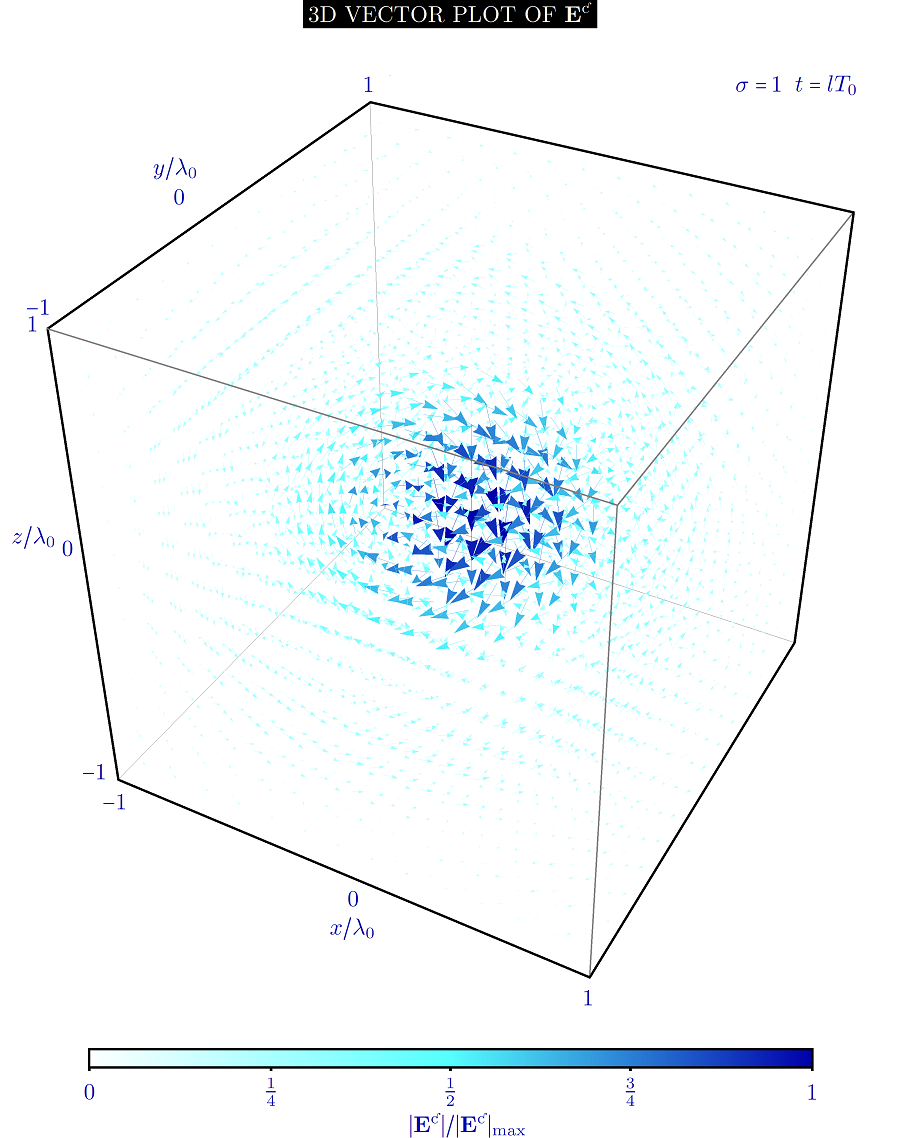}\\
\caption{\small The electric field $\mathbf{E}^\textrm{\textipa{\texthtc}}$ of the $\sigma=1$ form of of our electromagnetic tangle (\ref{Roses}), depicted as a three-dimensional vector plot at an instant of time: each blue arrow represents an electric field vector, colour-coded and scaled by magnitude.} 
\label{LRose3D}
\end{figure}

\begin{figure}[h!]
\centering
\includegraphics[width=\linewidth]{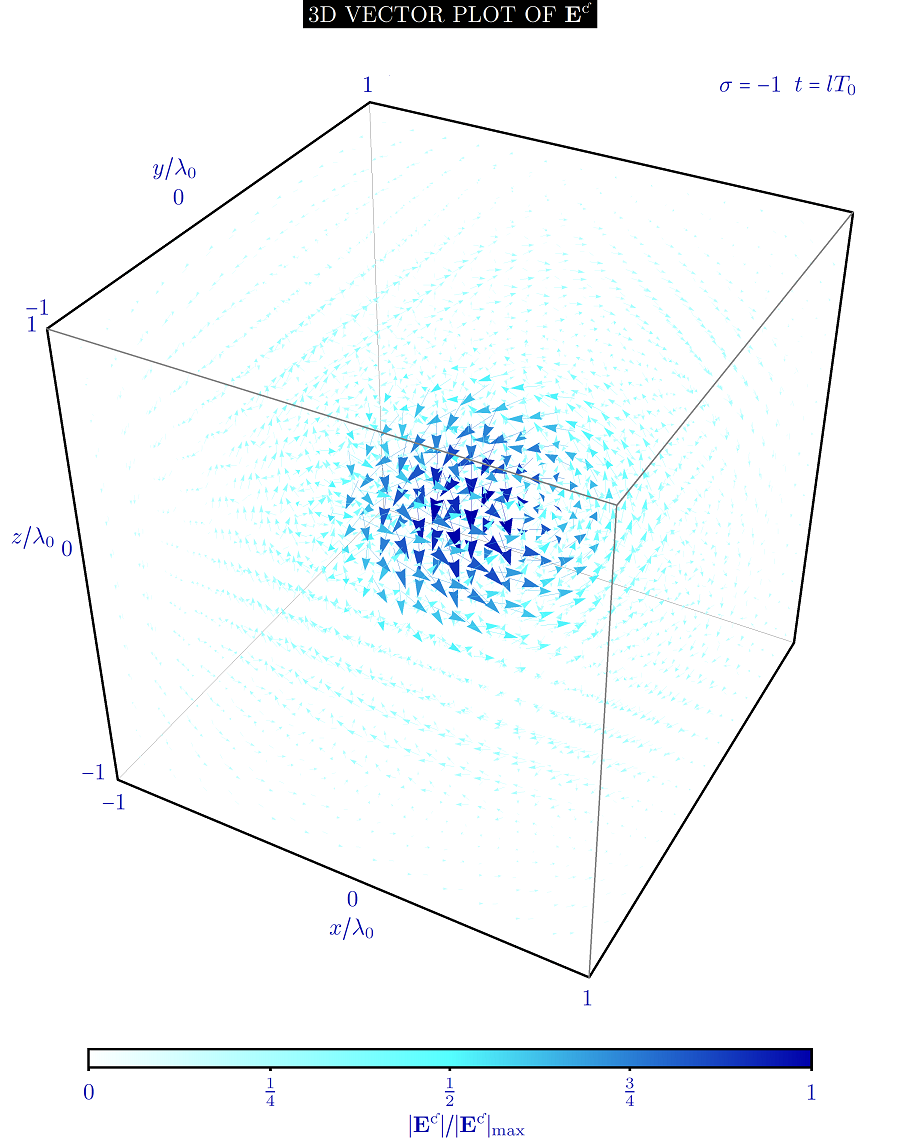}\\
\caption{\small The electric field $\mathbf{E}^\textrm{\textipa{\texthtc}}$ of the $\sigma=-1$ form of our electromagnetic tangle (\ref{Roses}), depicted as a three-dimensional vector plot at an instant of time: each blue arrow represents an electric field vector, colour-coded and scaled by magnitude.} 
\label{DRose3D}
\end{figure}

Taking $\tilde{\mathtt{A}}'_0=1/\sqrt{2}$ and $\tilde{\mathtt{B}}'_0=\textrm{i}\sigma/\sqrt{2}$ with $\sigma=\pm1$ in (\ref{UEDe1}) and (\ref{UEDb1}) corresponds to having each plane wave left- or right circularly polarised \cite{Jackson99,Fresnel32} and gives our `electromagnetic tangle':
\begin{eqnarray}
\mathbf{E}^\textrm{\textipa{\texthtc}}&=&\frac{E_0}{\sqrt{2}} (-\sigma\hat{\pmb{\phi}}f+\hat{\pmb{s}}g+\hat{\pmb{z}}h)\cos\left(\omega_0 t\right) \\
\mathbf{B}^\textrm{\textipa{\texthtc}}&=&\frac{E_0}{\sqrt{2}c}\left[\hat{\pmb{\phi}}f-\sigma\left(\hat{\pmb{s}}g+\hat{\pmb{z}}h\right)\right]\sin\left(\omega_0 t\right) ,
\end{eqnarray}
so-named because the electric and magnetic field lines encode a `tangle' of knots as described in more detail below. Our electromagnetic tangle is chiral \cite{Kelvin94}: $\sigma=1$ and $\sigma=-1$ give its enantiomorphs. The electric field $\mathbf{E}^\textrm{\textipa{\texthtc}}$ of the $\sigma=1$ form of our electromagnetic tangle is depicted in Fig. \ref{LRose3D} and that of the $\sigma=-1$ form is depicted in Fig. \ref{DRose3D}. Note that the magnetic field $\mathbf{B}^\textrm{\textipa{\texthtc}}$ of our electromagnetic tangle is proportional to a time-translated version of $\mathbf{E}^\textrm{\textipa{\texthtc}}$ in that $\mathbf{B}^\textrm{\textipa{\texthtc}}(t)=\sigma\mathbf{E}^\textrm{\textipa{\texthtc}}(t+T_0/4)/c$. Moreover, our electromagnetic tangle can be regarded as a superposition of our electric ring (section \ref{Rings}) and a time-translated version of our electric globule (section \ref{Globules}):
\begin{eqnarray}
\mathbf{E}^\textrm{\textipa{\texthtc}}(t)&=&\frac{1}{\sqrt{2}}\left[-\sigma\mathbf{E}^\textrm{\textipa{\textscripta}}(t)+\mathbf{E}^\textrm{\textipa{\texthtb}}(t+T_0/4)\right] \\
\mathbf{B}^\textrm{\textipa{\texthtc}}(t)&=&\frac{1}{\sqrt{2}}\left[-\sigma\mathbf{B}^\textrm{\textipa{\textscripta}}(t)+\mathbf{B}^\textrm{\textipa{\texthtb}}(t+T_0/4)\right],
\end{eqnarray}
as was suggested to the author by Stephen M Barnett and J\"{o}rg B G\"{o}tte. 

\begin{figure}[h!]
\centering
\includegraphics[width=\linewidth]{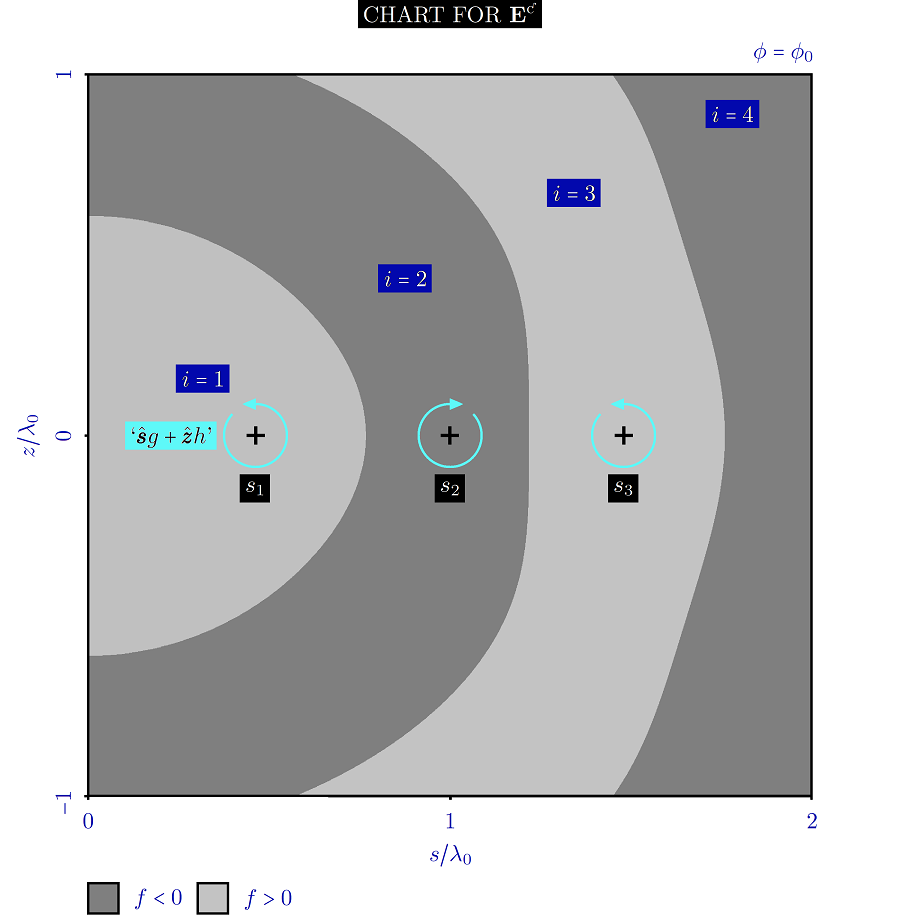}\\
\caption{\small A cross section of the $i=1,2,\dots$ tori defined by $f=0$, with regions for which $f<0$ in dark grey; regions for which $f>0$ in light grey; points for which $g(s=s_i,z=0)=h(s=s_i,0)=0$ indicated by black crosses and the sense of circulation of $\hat{\pmb{s}}g+\hat{\pmb{z}}h$ about these points indicated by light-blue arrows.} 
\label{Rose2D}
\end{figure}

The basic form of the electric field lines of our electromagnetic tangle, which are non-vanishing when $t\ne(1/2+l)T_0/2$, can be appreciated by considering the following, with reference to the chart in Fig. \ref{Rose2D}. In a plane with $\phi=\phi_0$, contours for which $f=0$ delimit crescent-shaped areas. The sign of $f$ is constant within the $i$th crescent ($i\in\{1,2,\dots\}$) and differs between the $i$th and $(i+1)$th crescent. Within the $i$th crescent there is one point a distance $s_i$ along the $s$ axis at $z=0$ about which the two-dimensional vector field $\hat{\pmb{s}}g+\hat{\pmb{z}}h$ circulates $(g(s=s_i,z=0)=h(s=s_i,z=0)=0)$, counter-clockwise if $f>0$ or clockwise if $f<0$. In three dimensions, the crescents define tori (obtained by revolving them about the $z$ axis). We now confine our attention within the $i$th such torus, without loss of generality: electric field lines do not pass between tori. There is one circular electric field line in the $z=0$ plane:
\begin{equation}
\mathbf{E}^\textrm{\textipa{\texthtc}}(s=s_i,z=0)=-\frac{1}{\sqrt{2}}\hat{\pmb{\phi}}\sigma f(s=s_i,z=0)E_0\cos\left(\omega_0 t\right), \label{unknot}
\end{equation}
which points in the $+\phi$ direction if $-\sigma f(s_i,0)\cos(\omega_0 t)>0$ or the $-\phi$ direction if $-\sigma f(s_i,0)\cos(\omega_0 t)<0$. The other electric field lines follow this line azimuthally (because $f$ has the same sign everywhere) whilst twisting around it (because of the circulation inherent to $\hat{\pmb{s}}g+\hat{\pmb{z}}h$) to form right-handed helices if $\sigma=1$ or left-handed helices if $\sigma=-1$. Analogous observations can be made for the magnetic field lines, of course, which are non-vanishing when $t\ne l T_0/2$. 

\begin{figure}[h!]
\centering
\includegraphics[width=0.95\linewidth]{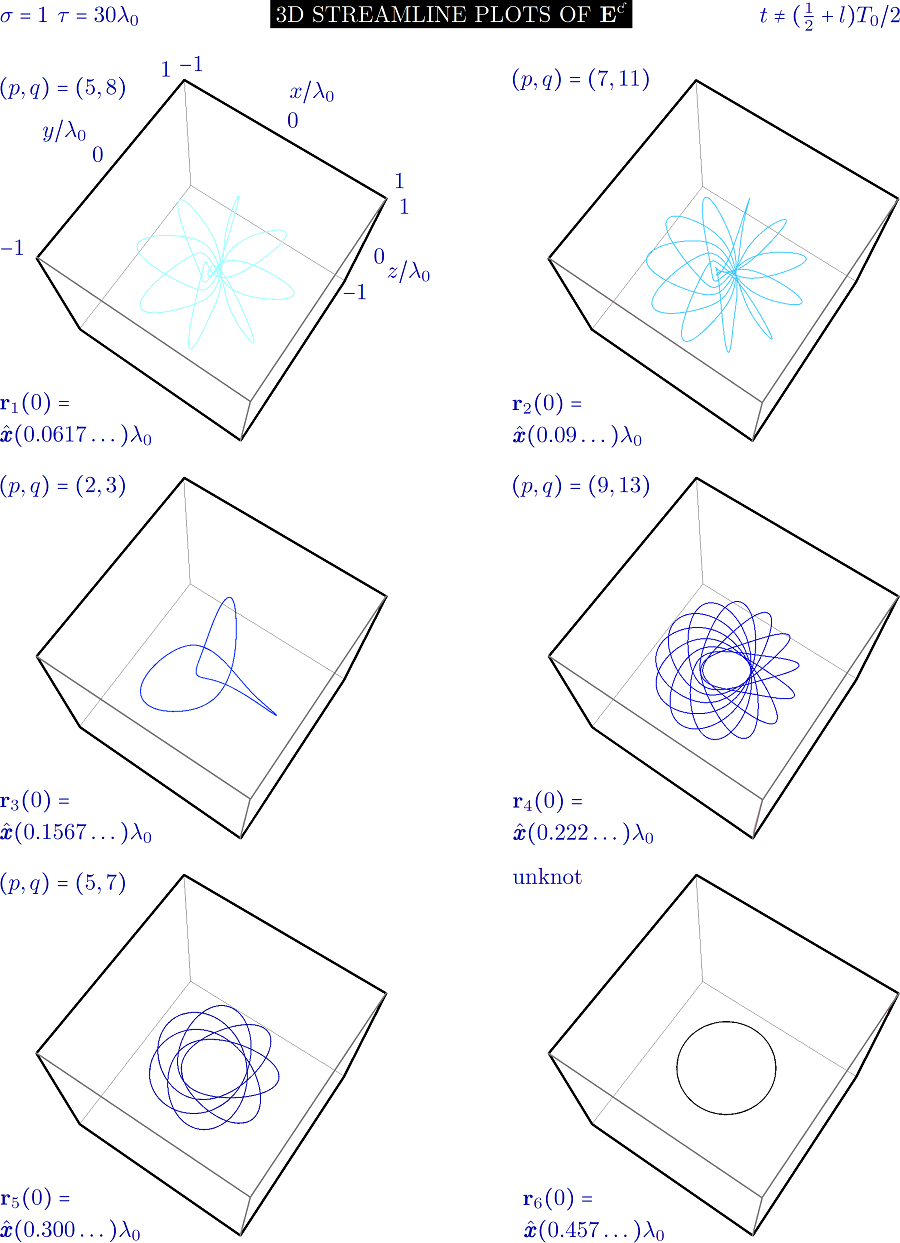}\\
\caption{\small A selection of torus-knotted electric field lines from the $i=1$ torus of the $\sigma=1$ form of our electromagnetic tangle (section \ref{Roses}) and their $(p,q)$ designations \cite{Livingstone93}, with each depicted as a three-dimensional streamline plot at an instant of time.} 
\label{RoseLines1}
\end{figure}

\begin{figure}[h!]
\centering
\includegraphics[width=0.95\linewidth]{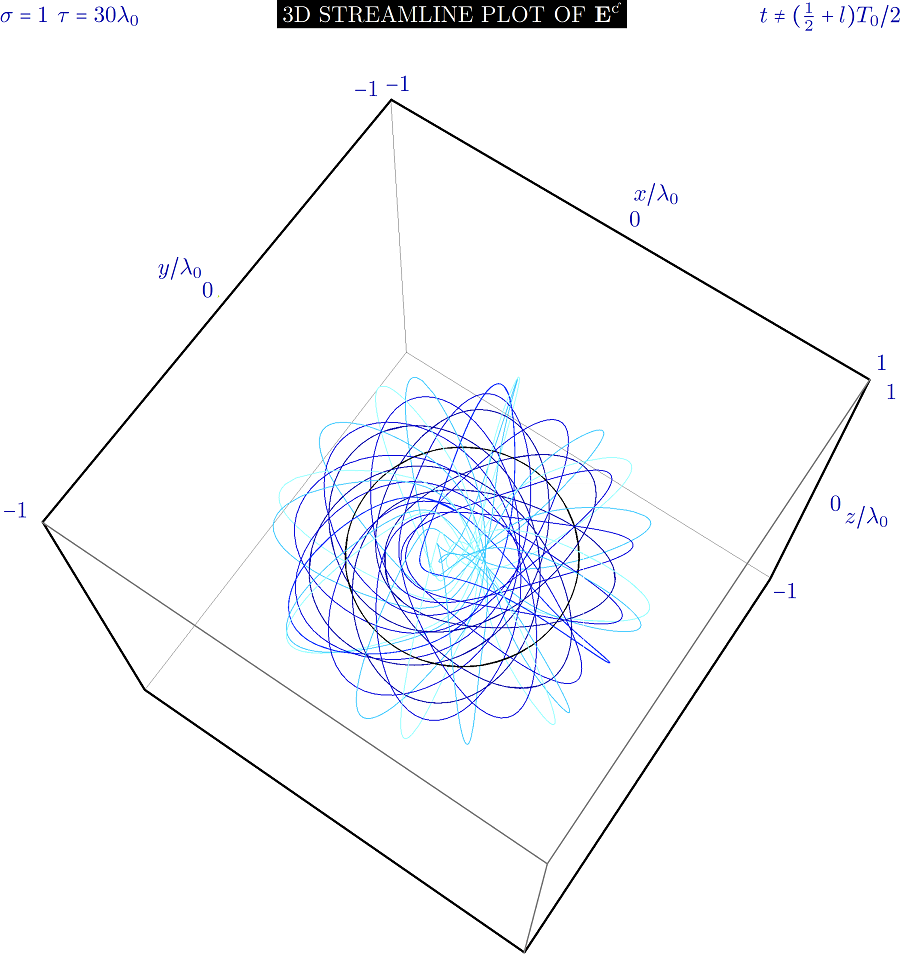}\\
\caption{\small The electric field lines depicted in Fig. \ref{RoseLines1}, shown together.} 
\label{RoseLines2}
\end{figure}

The electric field lines at any given time $t\ne(1/2+l)T_0/2$ can be examined by numerically integrating the streamline equation 
\begin{equation}
\frac{\textrm{d}\mathbf{r}(\tau)}{\textrm{d}\tau}=\frac{\mathbf{E}^\textrm{\textipa{\texthtc}}[\mathbf{r}(\tau)]}{|\mathbf{E}^\textrm{\textipa{\texthtc}}[\mathbf{r}(\tau)]|},
\end{equation}
giving the trajectory of the $j$th electric field line as
\begin{equation}
\mathbf{r}_j (\tau)=\mathbf{r}_j(0)+\int_0^\tau \frac{\mathbf{E}^\textrm{\textipa{\texthtc}}[\mathbf{r}(\tau')]}{|\mathbf{E}^\textrm{\textipa{\texthtc}}[\mathbf{r}(\tau')]|}\textrm{d}\tau', \label{streamline}
\end{equation}
where $\tau$ is an arc length which increases from $0$ as one follows the line from the seed position $\mathbf{r}_j(0)$. We find that certain electric field lines resemble familiar torus knots (with the circular electric field lines given by (\ref{unknot}) being unknots, of course) \cite{Livingstone93}. A selection of these lines are depicted separately in Fig. \ref{RoseLines1} and together in Fig. \ref{RoseLines2}. Again, analogous observations can be made for the magnetic field lines. 

Interestingly, there is a sense in which the chirality \cite{Kelvin94} of the helical field lines in our electromagnetic tangle is opposite to that of the plane waves that comprise the tangle: for $\sigma=\pm 1$ the field lines form right- or left-handed helices, as described above, whereas for each wave the tips of the field vectors trace out left- or right-handed cylindrical helices about the direction of propagation \cite{Takeda14}. It should be noted, however, that the field lines of a circularly polarised plane wave are, like those of a linearly polarised plane wave (section \ref{Introduction}), straight lines of indefinite extent.

Our electromagnetic tangle might be regarded as the first explicit example of a \textit{monochromatic} electromagnetic knot of toroidal character. Persistent \textit{polychromatic} electromagnetic knots of toroidal character are described in \cite{Kedia13}. Knotted threads of darkness within monochromatic electromagnetic fields are described in \cite{Berry01a, Leach04, Dennis10a,Dunlop17}, for example. Loosely speaking, our electromagnetic tangle might be regarded as a complimentary structure: here we have knotted `threads of brightness' (specifically, non-vanishing electric and magnetic field lines) rather than darkness.

\newpage

\newpage

\newpage

\newpage

\newpage


\newpage
\subsection{The second kind}
\label{The second kind}
In the present subsection we introduce our unusual electromagnetic disturbances of the second kind. Any one of these can be regarded as a superposition of translated and perhaps rotated versions of our unusual electromagnetic disturbances of the first kind (section \ref{The first kind}). 

Our unusual electromagnetic disturbances of the second kind are emphatically `new': electromagnetic disturbances akin to our electric loop (section \ref{Loops}), our electric link (section \ref{Links}), our electric lines (section \ref{Lines}), our electric knots (section \ref{REALKnots}) and our electromagnetic cloud (section \ref{Clouds}), for example, do not appear to have been described before. 

At the level of theory employed in the present paper, there is no obvious upper limit to the `size' of an unusual electromagnetic disturbance of the second kind: one is free to imagine an x-ray \cite{Rontgen95} electric loop large enough to encircle a star or an infrared \cite{Herschel80a, Herschel80b, Herschel80c} electromagnetic cloud as big as a house, for example.


\subsubsection{Construction}
\label{General recipe II}
We can construct a valid electromagnetic disturbance by superposing translated and perhaps rotated versions of our unusual electromagnetic disturbances of the first kind, as Maxwell's equations (\ref{DivE})-(\ref{Maxwell}) are linear, homogeneous and isotropic \cite{Bessel-Hagen21}.

To realise such a superposition and thus obtain our unusual electromagnetic disturbances of the second kind, let us introduce $N\ge 2$ auxiliary coordinate systems, with the $n$th system ($n\in\{1,\dots,N\}$) described by right-handed Cartesian coordinates $x_n$, $y_n$ and $z_n$ orientated such that the associated unit vectors $\hat{\pmb{x}}_n$, $\hat{\pmb{y}}_n$ and $\hat{\pmb{z}}_n$ are given in terms of Euler angles $\phi_n$, $\theta_n$ and $\chi_n$ as 
\begin{eqnarray}
\hat{\pmb{x}}_n&=&\hat{\pmb{x}}\ell^{(11)}_n+\hat{\pmb{y}}\ell^{(12)}_n+\hat{\pmb{z}}\ell^{(13)}_n, \\
\hat{\pmb{y}}_n&=&\hat{\pmb{x}}\ell^{(21)}_n+\hat{\pmb{y}}\ell^{(22)}_n+\hat{\pmb{z}}\ell^{(23)}_n \\
\hat{\pmb{z}}_n&=&\hat{\pmb{x}}\ell^{(31)}_n+\hat{\pmb{y}}\ell^{(32)}_n+\hat{\pmb{z}}\ell^{(33)}_n 
\end{eqnarray}
with the origin $x_n=y_n=z_n=0$ located at $x=X_n$, $y=Y_n$ and $z=Z_n$, where
\begin{eqnarray}
\ell^{(11)}_n&=&\cos\phi_n\cos\theta_n\cos\chi_n-\sin\phi_n\sin\chi_n, \\
\ell^{(12)}_n&=&-\sin\phi_n\cos\chi_n-\cos\phi_n\cos\theta_n\sin\chi_n, \\
\ell^{(13)}_n&=&\cos\phi_n\sin\theta_n, \\
\ell^{(21)}_n&=&\sin\phi_n\cos\theta_n\cos\chi_n+\cos\phi_n\sin\chi_n, \\
\ell^{(22)}_n&=&\cos\phi_n\cos\chi_n-\sin\phi_n\cos\theta_n\sin\chi_n, \\
\ell^{(23)}_n&=&\sin\phi_n\sin\theta_n, \\
\ell^{(31)}_n&=&-\sin\theta_n\cos\chi_n, \\
\ell^{(32)}_n&=&\sin\theta_n\sin\chi_n \\
\ell^{(33)}_n&=&\cos\theta_n
\end{eqnarray}
are direction cosines. It follows from the above together with (\ref{UEDe1}) and (\ref{UEDb1}) that our unusual electromagnetic disturbances of the second kind are given by
\begin{eqnarray}
\mathbf{E}&=&\Re \left\{\sum_{n=1}^NE_0\left[\textrm{i}\tilde{\mathtt{B}}'_{0n}\hat{\pmb{\phi}}_n f_n+\tilde{\mathtt{A}}'_{0n}\left(\hat{\pmb{s}}_n g_n+\hat{\pmb{z}}_n h_n\right)\right]\textrm{e}^{-\textrm{i}\omega_0 t}\right\} \label{ESECOND} \\
\mathbf{B}&=&\Re\left\{\sum_{n=1}^N\frac{E_0}{c}\left[\textrm{i}\tilde{\mathtt{A}}'_{0n}\hat{\pmb{\phi}}_n f_n-\tilde{\mathtt{B}}'_{0n}\left(\hat{\pmb{s}}_ng_n+\hat{\pmb{z}}_nh_n\right)\right]\textrm{e}^{-\textrm{i}\omega_0 t}\right\} \label{BSECOND}
\end{eqnarray}
with the modulating scalar fields
\begin{eqnarray}
f_n&=&\int_0^\pi \textrm{J}_1 (k_0\sin\vartheta s_n)\cos \left(k_0\cos\vartheta z_n\right)\sin\vartheta\textrm{d}\vartheta, \label{fintegraln} \\
g_n&=&-\int_0^\pi \textrm{J}_1 (k_0\sin\vartheta s_n)\sin \left(k_0\cos\vartheta z_n \right)  \sin\vartheta \cos\vartheta \textrm{d}\vartheta \label{gintegraln} \\
h_n&=&-\int_0^\pi \textrm{J}_0 (k_0\sin\vartheta s_n)\cos\left(k_0\cos\vartheta z_n \right)\sin^2\vartheta \textrm{d}\vartheta, \label{hintegraln}
\end{eqnarray}
where the $\tilde{\mathtt{A}}_{n0}'$ and $\tilde{\mathtt{B}}_{n0}'$ are complex constants and the $\phi_n'$, $s_n$ and $z_n$ are cylindrical coordinates defined such that
\begin{eqnarray}
x_n&=& s_n\cos\phi_n' \\
y_n&=& s_n\sin\phi_n'
\end{eqnarray}
with associated unit vectors 
\begin{eqnarray}
\hat{\pmb{\phi}}_n&=&-\hat{\pmb{x}}_n\sin\phi_n'+\hat{\pmb{y}}_n\cos\phi_n' \\
\hat{\pmb{s}}_n&=&\hat{\pmb{x}}_n\cos\phi_n'+\hat{\pmb{y}}_n\sin\phi_n'.
\end{eqnarray}
A particular unusual electromagnetic disturbance of the second kind is determined by specifying $N$, the $\phi_n$, the $\theta_n$, the $\chi_n$, the $X_n$, the $Y_n$, the $Z_n$, the $\tilde{\mathtt{A}}'_{0n}$ and the $\tilde{\mathtt{B}}'_{0n}$. We give explicit examples in section \ref{Loops}, section \ref{Links}, section \ref{Lines}, section \ref{REALKnots} and section \ref{Clouds}.

It is also possible to realise (\ref{ESECOND}) and (\ref{BSECOND}) via appropriate choices of $\tilde{\mathtt{A}}'$ and $\tilde{\mathtt{B}}'$ in (\ref{Emono}) and (\ref{Bmono}), of course. The forms required turn out to be particularly lengthy, however, and we refrain, therefore, from reproducing them explicitly in the present paper.


\newpage
\subsubsection{Electric loop}
\label{Loops}
An unusual electromagnetic disturbance in which some of the electric field lines form a particularly prominent loop-like feature (or features) at certain times can be constructed by superposing spatially translated and perhaps rotated versions of our electric ring (section \ref{Rings}). Various forms are possible.

\begin{figure}[h!]
\centering
\includegraphics[width=\linewidth]{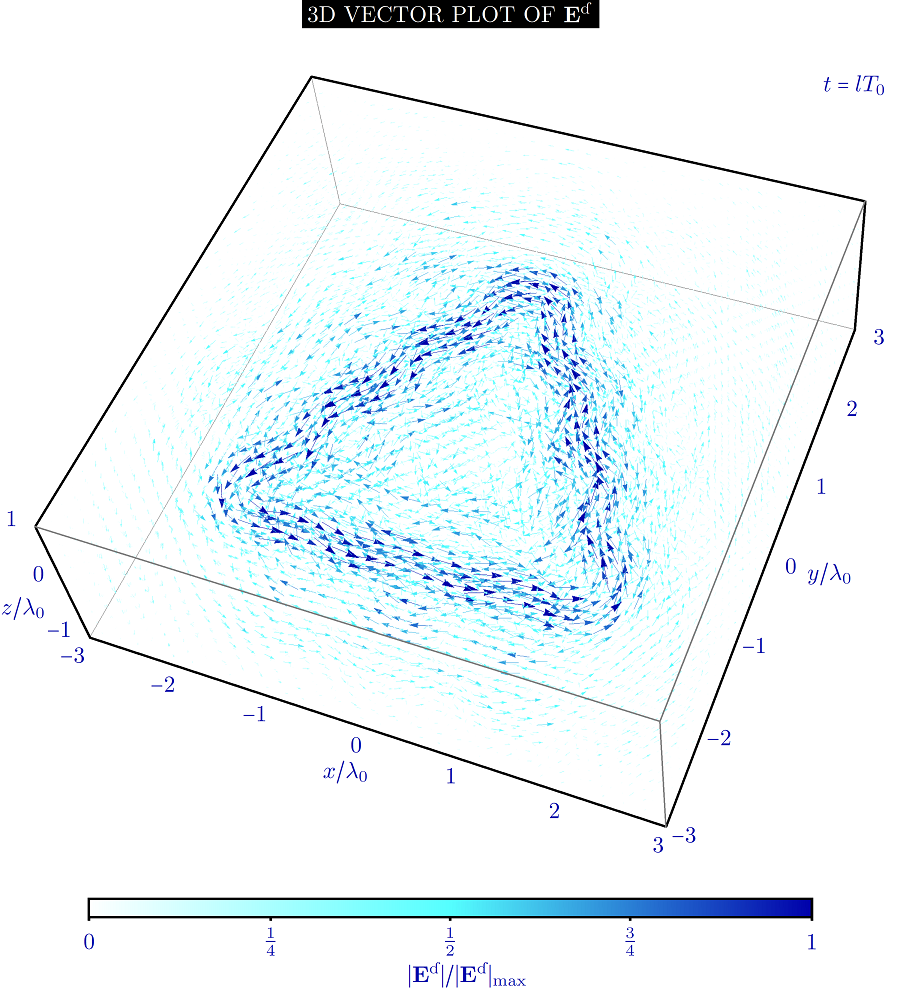}\\
\caption{\small The electric field $\mathbf{E}^\textrm{\textipa{\texthtd}}$ of our electric loop (section \ref{Loops}), depicted as a three-dimensional vector plot at an instant of time: each blue arrow represents an electric field vector, colour-coded and scaled by magnitude.} 
\label{Loop3D}
\end{figure}

Consider for example the triangular `electric loop'\footnote{In the present paper `electric loop' is not to be confused with `looped electric field line'.} specified by taking $N=21$ and
\begin{center}
\begin{tabular}{r r r r r r r r r}
$n$ & $\phi_n$ & $\theta_n$ & $\chi_n$ & $X_n/\lambda_0$ & $Y_n/\lambda_0$ & $Z_n/\lambda_0$  & $\tilde{\mathtt{A}}_{0n}'$  & $\tilde{\mathtt{B}}_{0n}'$ \\  
$1$ & $0$ & $0$ & $0$ & 0 & $5\sqrt{3}d/2$ & $0$  & $0$  & $-\textrm{i}$  \\  
$2$ & $0$ & $0$ & $0$ & $-d$ & $3\sqrt{3}d/2$ & $0$  & $0$  & $-\textrm{i}$  \\  
$3$ & $0$ & $0$ & $0$ & $d$ & $3\sqrt{3}d/2$ & $0$  & $0$  & $-\textrm{i}$  \\  
$4$ & $0$ & $0$ & $0$ & $-2d$ & $\sqrt{3}d/2$ & $0$  & $0$  &$-\textrm{i}$  \\  
$5$ & $0$ & $0$ & $0$ & $0$ & $\sqrt{3}d/2$ & $0$  & $0$  & $-\textrm{i}$  \\  
$6$ & $0$ & $0$ & $0$ & $2d$ & $\sqrt{3}d/2$ & $0$  & $0$  &$-\textrm{i}$  \\  
$7$ & $0$ & $0$ & $0$ & $-3d$ & $-\sqrt{3}d/2$ & $0$  & $0$  & $-\textrm{i}$  \\  
$8$ & $0$ & $0$ & $0$ & $-d$ & $-\sqrt{3}d/2$ & $0$  & $0$  & $-\textrm{i}$  \\  
$9$ & $0$ & $0$ & $0$ & $d$ & $-\sqrt{3}d/2$ & $0$  & $0$  & $-\textrm{i}$  \\  
$10$ & $0$ & $0$ & $0$ & $3d$ & $-\sqrt{3}d/2$ & $0$  & $0$  & $-\textrm{i}$  \\  
$11$ & $0$ & $0$ & $0$ & $-4d$ & $-3\sqrt{3}d/2$ & $0$  & $0$  & $-\textrm{i}$  \\  
$12$ & $0$ & $0$ & $0$ & $-2d$ & $-3\sqrt{3}d/2$ & $0$  & $0$  & $-\textrm{i}$  \\  
$13$ & $0$ & $0$ & $0$ & $0$ & $-3\sqrt{3}d/2$ & $0$  & $0$  & $-\textrm{i}$  \\  
$14$ & $0$ & $0$ & $0$ & $2d$ & $-3\sqrt{3}d/2$ & $0$  & $0$  &$-\textrm{i}$  \\  
$15$ & $0$ & $0$ & $0$ & $4d$ & $-3\sqrt{3}d/2$ & $0$  & $0$  & $-\textrm{i}$  \\  
$16$ & $0$ & $0$ & $0$ & $-5d$ & $-5\sqrt{3}d/2$ & $0$  & $0$  &$-\textrm{i}$  \\  
$17$ & $0$ & $0$ & $0$ & $-3d$ & $-5\sqrt{3}d/2$ & $0$  & $0$  & $-\textrm{i}$  \\  
$18$ & $0$ & $0$ & $0$ & $-d$ & $-5\sqrt{3}d/2$ & $0$  & $0$  & $-\textrm{i}$  \\  
$19$ & $0$ & $0$ & $0$ & $d$ & $-5\sqrt{3}d/2$ & $0$  & $0$  & $-\textrm{i}$  \\  
$20$ & $0$ & $0$ & $0$ & $3d$ & $-5\sqrt{3}d/2$ & $0$  & $0$  & $-\textrm{i}$  \\  
$21$ & $0$ & $0$ & $0$ & $5d$ & $-5\sqrt{3}d/2$ & $0$  & $0$  & $-\textrm{i}$  \\  
\end{tabular}
\end{center}
in (\ref{ESECOND}) and (\ref{BSECOND}), where it is to be understood that $d=0.33$: the area bounded by our electric loop is tiled with electric rings such that the electric fields of the rings largely cancel within the boundary of the loop whilst those on the loop connect. The formation of the loop is thus analogous to the formation of bound surface currents in magnetic media \cite{Griffiths99, Jackson99}. The electric field $\mathbf{E}^\textrm{\textipa{\texthtd}}$ of our electric loop is depicted in Fig. \ref{Loop3D}. 

Electric loops, including non-planar electric loops, can also be realised as closed electric lines (section \ref{Lines}).

\newpage


\newpage
\subsubsection{Electric link}
\label{Links}
An unusual electromagnetic disturbance in which some of the electric field lines form a particularly prominent link-like feature (or features) at certain times can be constructed by superposing electric loops (section \ref{Rings}). Various forms are possible.

\begin{figure}[h!]
\centering
\includegraphics[width=\linewidth]{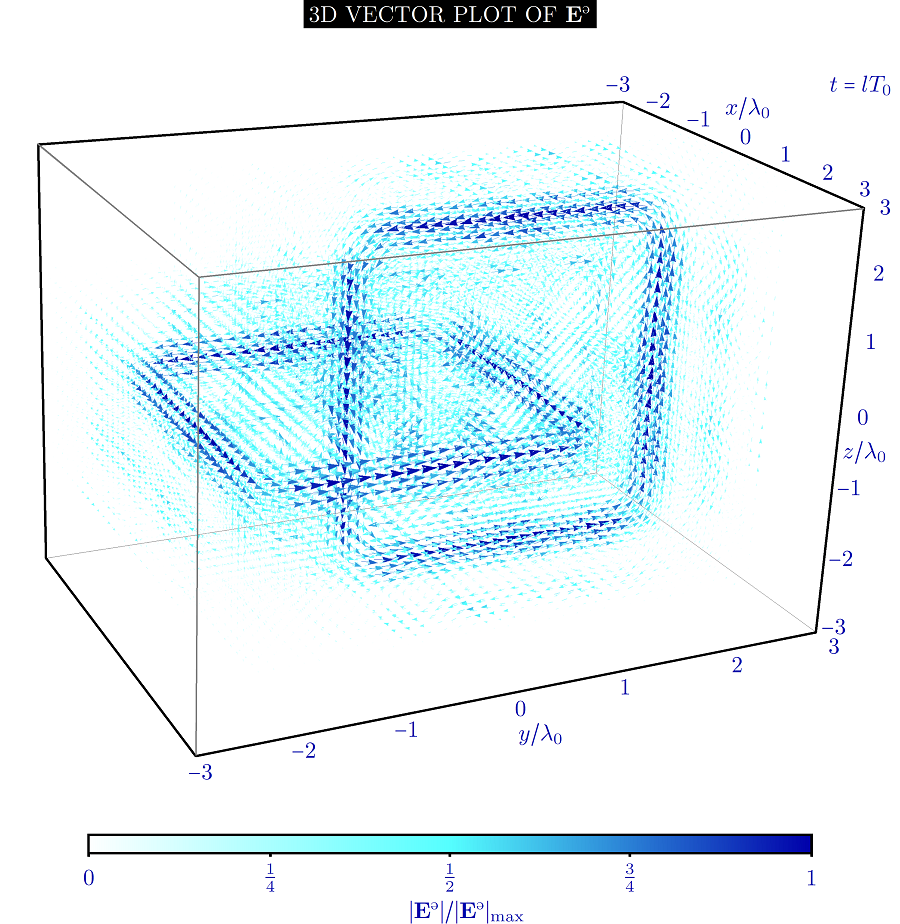} \\
\caption{\small The electric field $\mathbf{E}^\textrm{\textipa{\textschwa}}$ of our electric link (\ref{Links}), depicted as a three-dimensional vector plot at an instant of time: each blue arrow represents an electric field vector, colour-coded and scaled by magnitude. Each of the electric rings comprising the link has been plotted for $|z_n|\le\lambda_0$ only, for the sake of clarity.} 
\label{Link3D}
\end{figure}

Consider for example the `electric link' specified by taking $N=32$ and
\begin{center}
\begin{tabular}{r r r r r r r r r}
$n$ & $\phi_n$ & $\theta_n$ & $\chi_n$ & $X_n/\lambda_0$ & $Y_n/\lambda_0$ & $Z_n/\lambda_0$  & $\tilde{\mathtt{A}}_{0n}'$  & $\tilde{\mathtt{B}}_{0n}'$ \\  
$1$ & $0$ & $0$ & $0$ & $-3d/2$ & $-(3d+D)/2$ & $0$  & $0$  & $-\textrm{i}$  \\  
$2$ & $0$ & $0$ & $0$ & $-3d/2$ & $-(d+D)/2$ & $0$  & $0$  & $-\textrm{i}$  \\  
$3$ & $0$ & $0$ & $0$ & $-3d/2$ & $(d-D)/2$ & $0$  & $0$  & $-\textrm{i}$  \\  
$4$ & $0$ & $0$ & $0$ & $-3d/2$ & $(3d-D)/2$ & $0$  & $0$  & $-\textrm{i}$  \\   
$5$ & $0$ & $0$ & $0$ & $-d/2$ & $-(3d+D)/2$ & $0$  & $0$  & $-\textrm{i}$  \\  
$6$ & $0$ & $0$ & $0$ & $-d/2$ & $-(d+D)/2$ & $0$  & $0$  & $-\textrm{i}$  \\   
$7$ & $0$ & $0$ & $0$ & $-d/2$ & $(d-D)/2$ & $0$  & $0$  & $-\textrm{i}$  \\  
$8$ & $0$ & $0$ & $0$ & $-d/2$ & $(3d-D)/2$ & $0$  & $0$  & $-\textrm{i}$  \\  
$9$ & $0$ & $0$ & $0$ & $d/2$ & $-(3d+D)/2$ & $0$  & $0$  & $-\textrm{i}$  \\  
$10$ & $0$ & $0$ & $0$ & $d/2$ & $-(d+D)/2$ & $0$  & $0$  & $-\textrm{i}$  \\  
$11$ & $0$ & $0$ & $0$ & $d/2$ & $(d-D)/2$ & $0$  & $0$  & $-\textrm{i}$  \\  
$12$ & $0$ & $0$ & $0$ & $d/2$ & $(3d-D)/2$ & $0$  & $0$  & $-\textrm{i}$  \\  
$13$ & $0$ & $0$ & $0$ & $3d/2$ & $-(3d+D)/2$ & $0$  & $0$  & $-\textrm{i}$  \\  
$14$ & $0$ & $0$ & $0$ & $3d/2$ & $-(d+D)/2$ & $0$  & $0$  & $-\textrm{i}$  \\  
$15$ & $0$ & $0$ & $0$ & $3d/2$ & $(d-D)/2$ & $0$  & $0$  & $-\textrm{i}$  \\  
$16$ & $0$ & $0$ & $0$ & $3d/2$ & $(3d-D)/2$ & $0$  & $0$  & $-\textrm{i}$  \\    
$17$ & $0$ & $\pi/2$ & $0$ & $0$ & $-(3d-D)/2$ & $-3d/2$  & $0$  & $-\textrm{i}$  \\  
$18$ & $0$ & $\pi/2$ & $0$ & $0$ & $-(3d-D)/2$ & $-d/2$    & $0$  & $-\textrm{i}$  \\  
$19$ & $0$ & $\pi/2$ & $0$ & $0$ & $-(3d-D)/2$ & $d/2$    & $0$  & $-\textrm{i}$  \\  
$20$ & $0$ & $\pi/2$ & $0$ & $0$ & $-(3d-D)/2$ & $3d/2$    & $0$  & $-\textrm{i}$  \\  
$21$ & $0$ & $\pi/2$ & $0$ & $0$ & $-(d-D)/2$ & $-3d/2$   & $0$  & $-\textrm{i}$  \\  
$22$ & $0$ & $\pi/2$ & $0$ & $0$ & $-(d-D)/2$ & $-d/2$   & $0$  & $-\textrm{i}$  \\  
$23$ & $0$ & $\pi/2$ & $0$ &$0$ &  $-(d-D)/2$ & $d/2$   & $0$  & $-\textrm{i}$  \\  
$24$ & $0$ & $\pi/2$ & $0$ &$0$ &  $-(d-D)/2$ & $3d/2$  & $0$  & $-\textrm{i}$  \\  
$25$ & $0$ & $\pi/2$ & $0$ & $0$ & $(d+D)/2$ & $-3d/2$  & $0$  & $-\textrm{i}$  \\  
$26$ & $0$ & $\pi/2$ & $0$ & $0$ & $(d+D)/2$ & $-d/2$ & $0$  & $-\textrm{i}$  \\  
$27$ & $0$ & $\pi/2$ & $0$ & $0$ & $(d+D)/2$ & $d/2$ & $0$  & $-\textrm{i}$  \\  
$28$ & $0$ & $\pi/2$ & $0$ & $0$ & $(d+D)/2$ & $3d/2$  & $0$  & $-\textrm{i}$  \\  
$29$ & $0$ & $\pi/2$ & $0$ & $0$ & $(3d+D)/2$ & $-3d/2$  & $0$  & $-\textrm{i}$  \\  
$30$ & $0$ & $\pi/2$ & $0$ & $0$ & $(3d+D)/2$ & $-d/2$   & $0$  & $-\textrm{i}$  \\  
$31$ & $0$ & $\pi/2$ & $0$ & $0$ & $(3d+D)/2$ & $d/2$  & $0$  & $-\textrm{i}$  \\  
$32$ & $0$ & $\pi/2$ & $0$ & $0$ & $(3d+D)/2$ & $3d/2$   & $0$ & $-\textrm{i}$  \\   
\end{tabular}
\end{center}
in (\ref{ESECOND}) and (\ref{BSECOND}), where it is to be understood that $d=0.75$ and $D=1.50$: our electric link is comprised of two square electric loops, suitably orientated and positioned. The electric field $\mathbf{E}^\textrm{\textipa{\textschwa}}$ of our electric link is depicted in Fig. \ref{Link3D}. 

Our electric link might be regarded as another explicit example of a \textit{monochromatic} electromagnetic `knot'.

\newpage
\subsubsection{Electric lines}
\label{Lines}
An unusual electromagnetic disturbance in which some of the electric field lines form a particularly prominent line-like feature (or features) at certain times can be constructed by superposing spatially translated and perhaps rotated versions of our electric globule (section \ref{Globules}). Various forms are possible.

\begin{figure}[h!]
\centering
\includegraphics[width=0.9\linewidth]{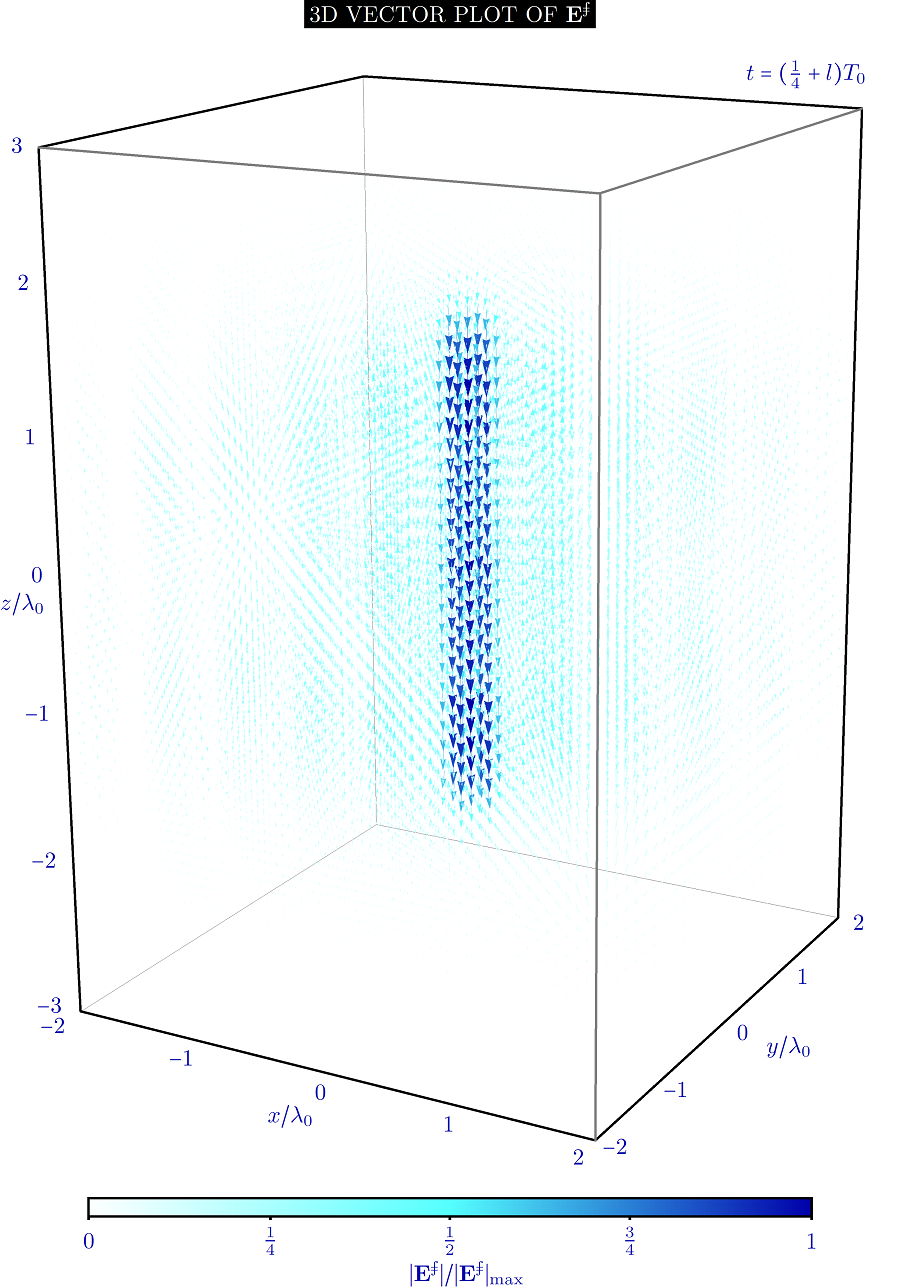}\\
\caption{\small The electric field $\mathbf{E}^\textrm{\textipa{\textdoublebaresh}}$ of our straight electric line (section \ref{Lines}), depicted as a three-dimensional vector plot at an instant of time: each blue arrow represents an electric field vector, colour-coded and scaled by magnitude.} 
\label{Rod3D}
\end{figure}

As our first example consider the `straight electric line'\footnote{In the present paper `electric line' is not to be confused with `electric field line'.} specified by taking $N=15$ and
\begin{center}
\begin{tabular}{r r r r r r r r r}
$n$ & $\phi_n$ & $\theta_n$ & $\chi_n$ & $X_n$ & $Y_n$ & $Z_n$  & $\tilde{\mathtt{A}}_{0n}'$  & $\tilde{\mathtt{B}}_{0n}'$ \\  
$1$ & $0$ & $0$ & $0$ & $0$ & $0$ & $-7d$  & $\textrm{i}$  & $0$  \\
$2$ & $0$ & $0$ & $0$ & $0$ & $0$ & $-6d$  & $\textrm{i}$  & $0$  \\  
$3$ & $0$ & $0$ & $0$ & $0$ & $0$ & $-5d$  & $\textrm{i}$  & $0$  \\  
$4$ & $0$ & $0$ & $0$ & $0$ & $0$ & $-4d$  & $\textrm{i}$  & $0$  \\  
$5$ & $0$ & $0$ & $0$ & $0$ & $0$ & $-3d$  & $\textrm{i}$  & $0$  \\  
$6$ & $0$ & $0$ & $0$ & $0$ & $0$ & $-2d$  & $\textrm{i}$  & $0$  \\  
$7$ & $0$ & $0$ & $0$ & $0$ & $0$ & $-d$  & $\textrm{i}$  & $0$  \\  
$8$ & $0$ & $0$ & $0$ & $0$ & $0$ & $0$  & $\textrm{i}$  & $0$  \\  
$9$ & $0$ & $0$ & $0$ & $0$ & $0$ & $d$  & $\textrm{i}$  & $0$  \\  
$10$ & $0$ & $0$ & $0$ & $0$ & $0$ & $2d$  & $\textrm{i}$  & $0$  \\  
$11$ & $0$ & $0$ & $0$ & $0$ & $0$ & $3d$  & $\textrm{i}$  & $0$  \\  
$12$ & $0$ & $0$ & $0$ & $0$ & $0$ & $4d$  & $\textrm{i}$  & $0$  \\  
$13$ & $0$ & $0$ & $0$ & $0$ & $0$ & $5d$  & $\textrm{i}$  & $0$  \\  
$14$ & $0$ & $0$ & $0$ & $0$ & $0$ & $6d$  & $\textrm{i}$  & $0$  \\  
$15$ & $0$ & $0$ & $0$ & $0$ & $0$ & $7d$  & $\textrm{i}$  & $0$  \\  
\end{tabular}
\end{center}
in (\ref{ESECOND}) and (\ref{BSECOND}), where it is to be understood that $d=0.25$: spatially translated versions of our electric globule are equally spaced along the $z$ axis like totems in a totem pole such that the straight electric line which results has a continuous appearance. The electric field $\mathbf{E}^\textrm{\textipa{\textdoublebaresh}}$ of our straight electric line is depicted in Fig. \ref{Rod3D}.

\newpage

\begin{figure}[h!]
\centering
\includegraphics[width=0.9\linewidth]{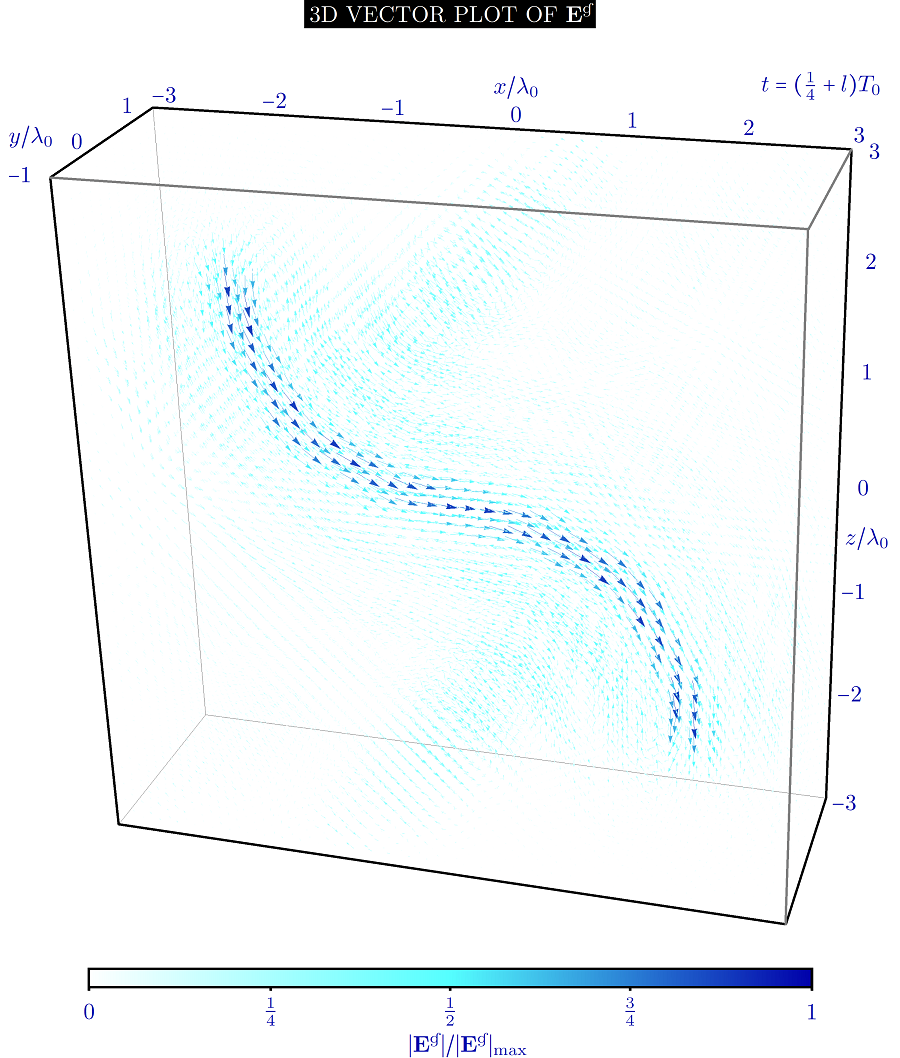}\\
\caption{\small The electric field $\mathbf{E}^\textrm{\textipa{\texthtg}}$ of our curved electric line (section \ref{Lines}), depicted as a three-dimensional vector plot at an instant of time: each blue arrow represents an electric field vector, colour-coded and scaled by magnitude.} 
\label{Line3D}
\end{figure}

\newpage
As our second example consider the `curved electric line' specified by taking $N=15$ and
\begin{center}
\begin{tabular}{r r r r r r r r r}
$n$ & $\phi_n$ & $\theta_n$ & $\chi_n$ & $X_n/\lambda_0$ & $Y_n/\lambda_0$ & $Z_n/\lambda_0$  & $\tilde{\mathtt{A}}_{0n}'$  & $\tilde{\mathtt{B}}_{0n}'$ \\  
$1$ & $\pi$ & $0$ & $0$ & $d\cos\theta_1$ & $0$ & $d(\sin\theta_{1}-1)$  & $\textrm{i}$  & $0$  \\  
$2$ & $\pi$ & $\pi/14$ & $0$ & $d\cos\theta_2$ & $0$ & $d(\sin\theta_2-1)$  &  $\textrm{i}$  & $0$  \\  
$3$ & $\pi$ & $\pi/7$ & $0$ & $d\cos\theta_3$ & $0$ & $d(\sin\theta_3-1)$  & $\textrm{i}$  & $0$  \\  
$4$ & $\pi$ & $3\pi/14$ & $0$ & $d\cos\theta_4$ & $0$ & $d(\sin\theta_4-1)$  &  $\textrm{i}$  & $0$  \\  
$5$ & $\pi$ & $2\pi/7$ & $0$ & $d\cos\theta_5$ & $0$ & $d(\sin\theta_5-1)$  &  $\textrm{i}$  & $0$  \\  
$6$ & $\pi$ & $5\pi/14$ & $0$ & $d\cos\theta_6$ & $0$ & $d(\sin\theta_6-1)$  &  $\textrm{i}$  & $0$  \\  
$7$ & $\pi$ & $3\pi/7$ & $0$ & $d\cos\theta_7$ & $0$ & $d(\sin\theta_7-1)$  &  $\textrm{i}$  & $0$  \\  
$8$ & $\pi$ & $\pi/2$ & $0$ & $d\cos\theta_8$ & $0$ & $d(\sin\theta_8-1)$  &  $\textrm{i}$  & $0$  \\  
$9$ & $\pi$ & $\pi$ & $0$ & $d\cos\theta_9$ & $0$ & $d(\sin\theta_9+1)$  &  $\textrm{i}$  & $0$  \\  
$10$ & $\pi$ & $15\pi/14$ & $0$ & $d\cos\theta_{10}$ & $0$ & $d(\sin\theta_{10}+1)$  &  $\textrm{i}$  & $0$  \\  
$11$ & $\pi$ & $8\pi/7$ & $0$ & $d\cos\theta_{11}$ & $0$ & $d(\sin\theta_{11}+1)$  &  $\textrm{i}$  & $0$  \\  
$12$ & $\pi$ & $17\pi/14$ & $0$ & $d\cos\theta_{12}$ & $0$ & $d(\sin\theta_{12}+1)$  &  $\textrm{i}$  & $0$  \\  
$13$ & $\pi$ & $9\pi/7$ & $0$ & $d\cos\theta_{13}$ & $0$ & $d(\sin\theta_{13}+1)$  &  $\textrm{i}$  & $0$  \\  
$14$ & $\pi$ & $19\pi/14$ & $0$ & $d\cos\theta_{14}$ & $0$ & $d(\sin\theta_{14}+1)$  &  $\textrm{i}$  & $0$  \\  
$15$ & $\pi$ & $10\pi/7$ & $0$ & $d\cos\theta_{15}$ & $0$ & $d(\sin\theta_{15}+1)$  &  $\textrm{i}$  & $0$  \\  
\end{tabular}
\end{center}
in (\ref{ESECOND}) and (\ref{BSECOND}), where it is to be understood that $d=2.00$: an S-shaped curve underlies our curved electric line and is populated with electric globules much as a string might be populated with beads, the globules being orientated and translated such that the electric field vectors at the origin of each globule in isolation are tangential with the underlying curve and the curved electric line which results has a continuous appearance. The electric field $\mathbf{E}^\textrm{\textipa{\texthtg}}$ of our curved electric line is depicted in Fig. \ref{Line3D}. 

\newpage


\newpage
\subsubsection{Electric knots}
\label{REALKnots}
An unusual electromagnetic disturbance in which some of the electric field lines form a particularly prominent feature resembling a torus or non-torus knot \cite{Livingstone93} at certain times can be realised as a closed electric line (section \ref{Lines}) with the appropriate topology. Various forms are possible.

\begin{figure}[h!]
\centering
\includegraphics[width=0.9\linewidth]{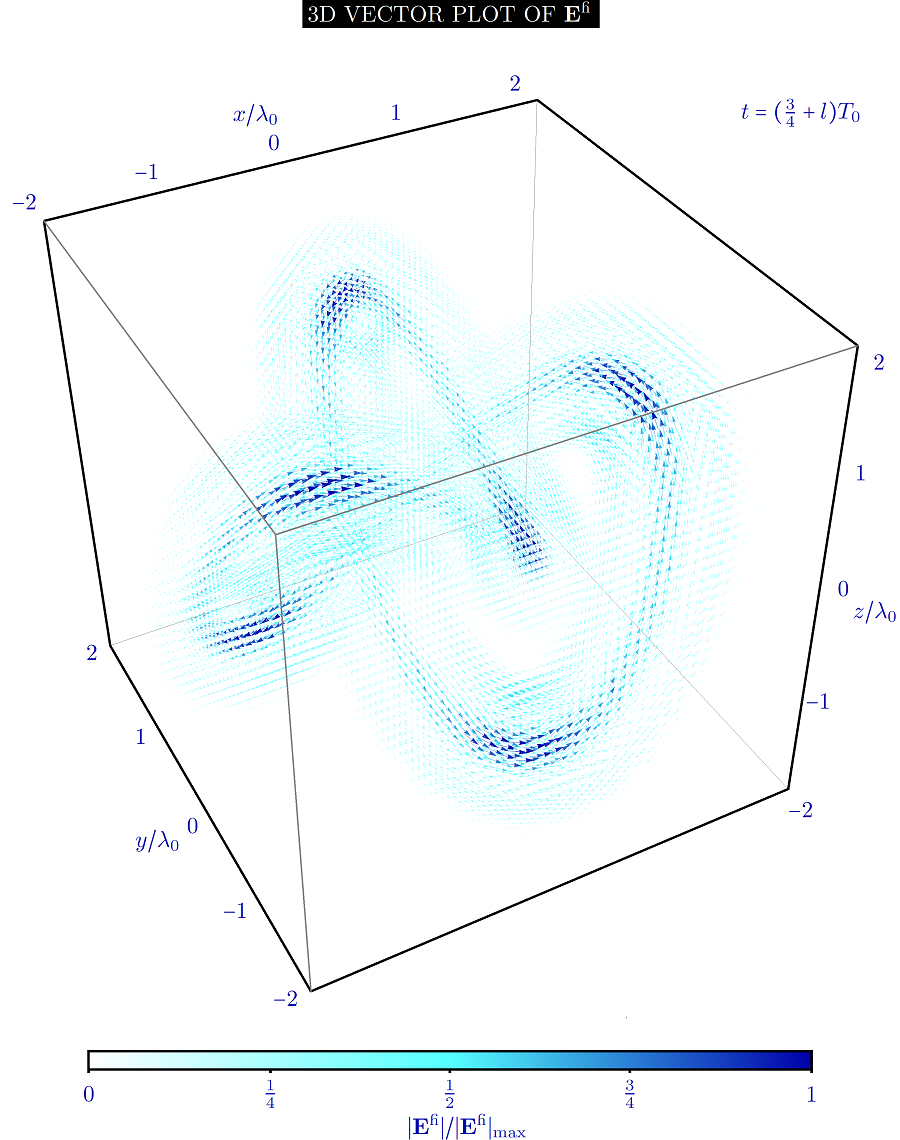} \\
\caption{\small The electric field $\mathbf{E}^\textrm{\textipa{\texthth}}$ of our electric torus knot (section \ref{REALKnots}), depicted as a three-dimensional vector plot at an instant of time: each blue arrow represents an electric field vector, colour-coded and scaled by magnitude. Each of the electric rings comprising the knot has been plotted for $|s_n|\le\lambda_0$ and $|z_n|\le\lambda_0$ only, for the sake of clarity and to render the task numerically tractable.} 
\label{Trefoil3D}
\end{figure}

As our first example consider the `electric torus knot' specified by taking $N=80$ and 
\begin{eqnarray}
\phi_n&=&\textrm{atan}2(\hat{v}_y,\hat{v}_x), \nonumber \\
\theta_n&=&\arccos\hat{v}_z,  \nonumber \\
\chi_n&=&0,  \nonumber \\
X'_n&=&d[\sin\mu_n+2\sin(2\mu_n)],  \nonumber \\
Y'_n&=& d[\cos \mu_n-2\cos(2 \mu_n )],  \nonumber \\
Z'_n&=&-2 d\sin(3\mu_n), \nonumber \\
\tilde{\mathtt{A}}_n'&=&\textrm{i}  \nonumber \\
\tilde{\mathtt{B}}_n'&=& 0, \nonumber 
\end{eqnarray}
where $\textrm{atan}2(Y,X)$ is the four-quadrant inverse tangent function, $\mu_n=2\pi(n-1)/N$ and it is to be understood that
\begin{eqnarray}
\hat{v}_x&=&\frac{\cos\mu_n+4 \cos(2\mu_n)}{\sqrt{35+8\cos(3\mu_n)+18 \cos(6\mu_n)}},  \nonumber \\
\hat{v}_y&=&\frac{\{-\sin\mu_n+4 \sin(2\mu_n)\}}{\sqrt{35+8\cos(3\mu_n)+18 \cos(6\mu_n)}},   \nonumber \\
\hat{v}_z&=&-\frac{6\cos(3\mu_n)}{\sqrt{35+8\cos(3\mu_n)+18 \cos(6\mu_n)}}   \nonumber 
\end{eqnarray}
and $d=0.90$: our electric torus knot is based upon a $(p,q)=(2,-3)$ torus knot, also known as a left-handed trefoil knot \cite{Livingstone93}. The electric field $\mathbf{E}^\textrm{\textipa{\texthth}}$ of our electric torus knot is depicted in Fig. \ref{Trefoil3D}. 

Our electric torus knot might be regarded as another explicit example of a \textit{monochromatic} electromagnetic knot of toroidal character, following our electromagnetic tangle (section \ref{Roses}). 

\newpage

\begin{figure}[h!]
\centering
\includegraphics[width=0.9\linewidth]{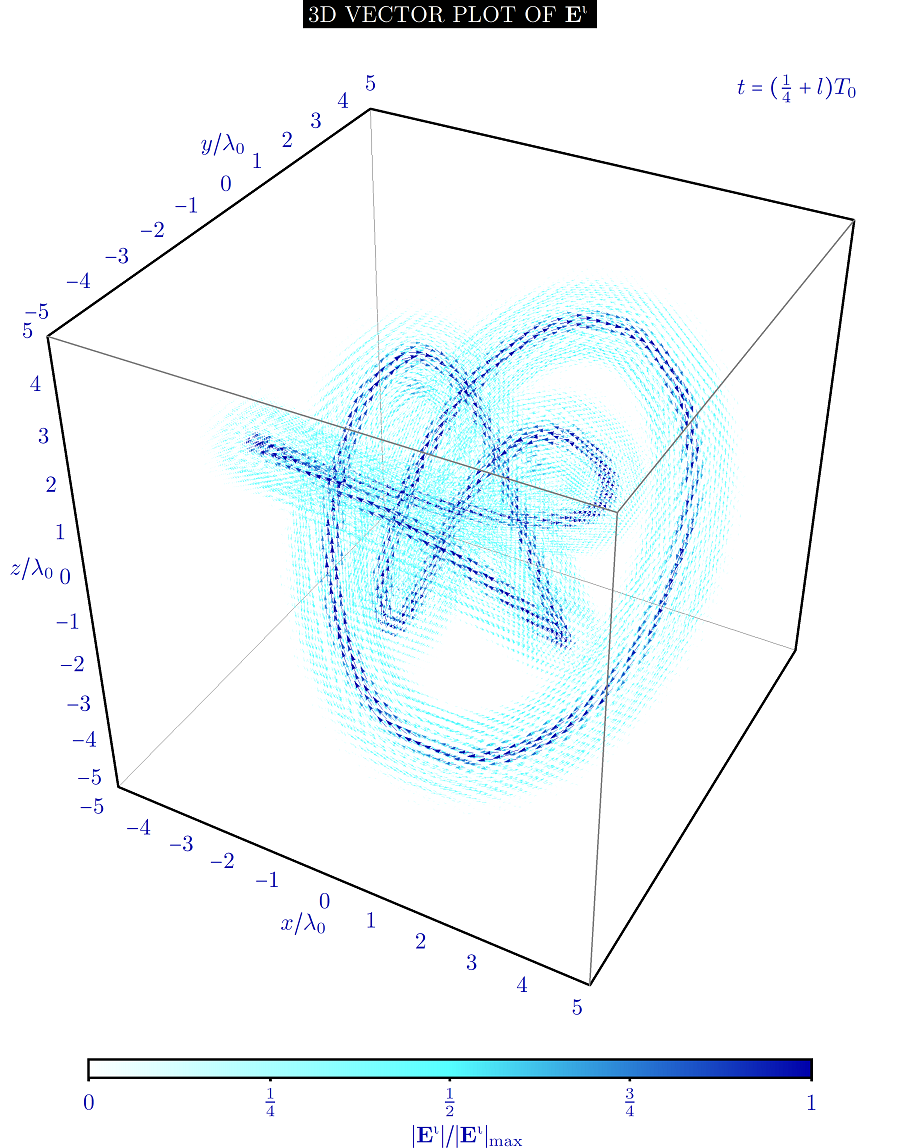}\\
\caption{\small The electric field $\mathbf{E}^\textrm{\textipa{\textiota}}$ of our non-torus electric knot (section \ref{REALKnots}), depicted as a three-dimensional vector plot at an instant of time: each blue arrow represents an electric field vector, colour-coded and scaled by magnitude. Each of the electric rings comprising the knot has been plotted for $|s_n|\le\lambda_0$ and $|z_n|\le\lambda_0$ only, for the sake of clarity and to render the task numerically tractable.} 
\label{8Knot3D}
\end{figure}

\newpage
As our second example consider the electric non-torus knot specified by taking $N=200$ and
\begin{eqnarray}
\phi_n&=&\textrm{atan}2(\hat{v}_y,\hat{v}_x), \nonumber \\
\theta_n&=&\arccos\hat{v}_z,  \nonumber \\
\chi_n&=&0,  \nonumber \\
X'_n&=&d[2+\cos(2\mu_n)]\cos(3\mu_n), \nonumber \\
Y'_n&=&d[2+\cos(2\mu_n)]\sin(3\mu_n), \nonumber \\
Z'_n&=&2d\sin(4\mu_n), \nonumber \\
\tilde{\mathtt{A}}_n'&=&\alpha_n\textrm{i} \nonumber \\
\tilde{\mathtt{B}}_n'&=&0 \nonumber
\end{eqnarray}
with the amplitudes $\alpha_n$ as described below, where it is to be understood that
\begin{eqnarray}
\hat{v}_x&=&\frac{-2\sin(2\mu_n)\cos(3\mu_n)-3[2+\cos(2\mu_n)]\sin(3\mu_n)}{\sqrt{[149+72\cos(2\mu_n)+5\cos(4\mu_n)+64\cos(8\mu_n)]/2}}, \nonumber \\
\hat{v}_y&=&\frac{-2\sin(2\mu_n)\sin(3\mu_n)+3[2+\cos(2\mu_n)]\cos(3\mu_n)}{\sqrt{[149+72\cos(2\mu_n)+5\cos(4\mu_n)+64\cos(8\mu_n)]/2}}, \nonumber \\
\hat{v}_z&=&\frac{8\cos(4\mu_n)}{\sqrt{[149+72\cos(2\mu_n)+5\cos(4\mu_n)+64\cos(8\mu_n)]/2}} \nonumber
\end{eqnarray}
and $d=1.33$: our electric non-torus knot is based upon a figure-of-eight knot \cite{Livingstone93}. The electric field $\mathbf{E}^\textrm{\textipa{\textiota}}$ of our electric non-torus knot is depicted in Fig. \ref{8Knot3D}. The $\alpha_n$ are obtained by first calculating $\mathbf{E}^\textrm{\textipa{\textiota}}$ with the $\alpha_n=1$, then taking $\alpha_n=E_0/|\mathbf{E}^\textrm{\textipa{\textiota}}(x=X_n,y=Y_n,z=Z_n,t\ne lT_0/2)|$ for the final $\mathbf{E}^\textrm{\textipa{\textiota}}$. This iteration gives the final electric non-torus knot a `cleaner' form than is obtained with the $\alpha_n=1$.

Our electric non-torus knot might be regarded as the first explicit example of a \textit{monochromatic} electromagnetic knot of non-toroidal character. A method by which to construct \textit{polychromatic} electromagnetic knots of non-toroidal character (at a particular time) is described in \cite{Kedia16}.

We note here once more, as in \S\ref{Roses}, that knotted threads of darkness within monochromatic electromagnetic fields are described in \cite{Berry01a, Leach04, Dennis10a,Dunlop17}, for example. Again, loosely speaking, our electric knots might be regarded as complimentary structures: here we have knotted `threads of brightness' (specifically, non-vanishing electric field lines) rather than darkness.

\newpage


\newpage
\subsubsection{Electromagnetic cloud}
\label{Clouds}
Random choices of the parameters defining our unusual electromagnetic disturbances of the second kind can yield unusual electromagnetic disturbances with prominent electric (and magnetic) features that resemble clouds. 

\begin{figure}[h!]
\centering
\includegraphics[width=\linewidth]{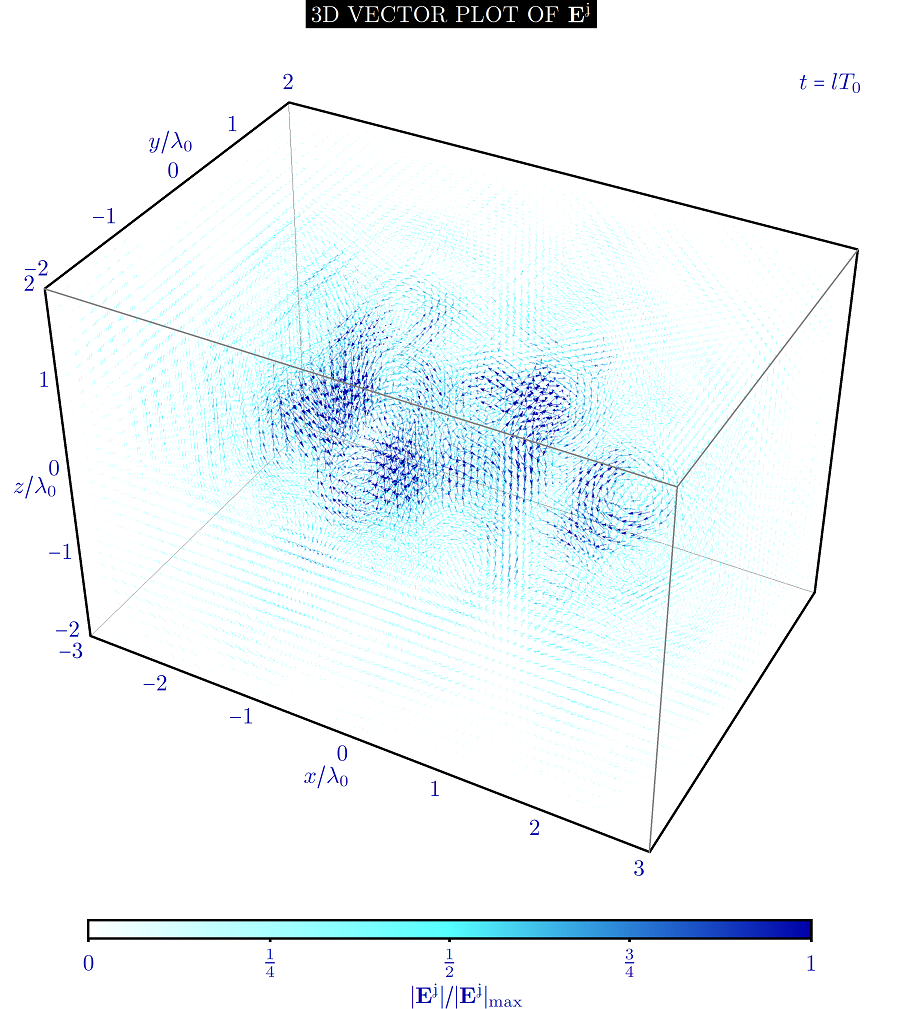}\\
\caption{\small The electric field $\mathbf{E}^\textrm{\textipa{\textctj}}$ of our electromagnetic cloud (section \ref{Clouds}) for $t=l T_0$, depicted as a three-dimensional vector plot: each blue arrow represents an electric field vector, colour-coded and scaled by magnitude.} 
\label{Cloud3Dt0}
\end{figure}

\begin{figure}[h!]
\centering
\includegraphics[width=\linewidth]{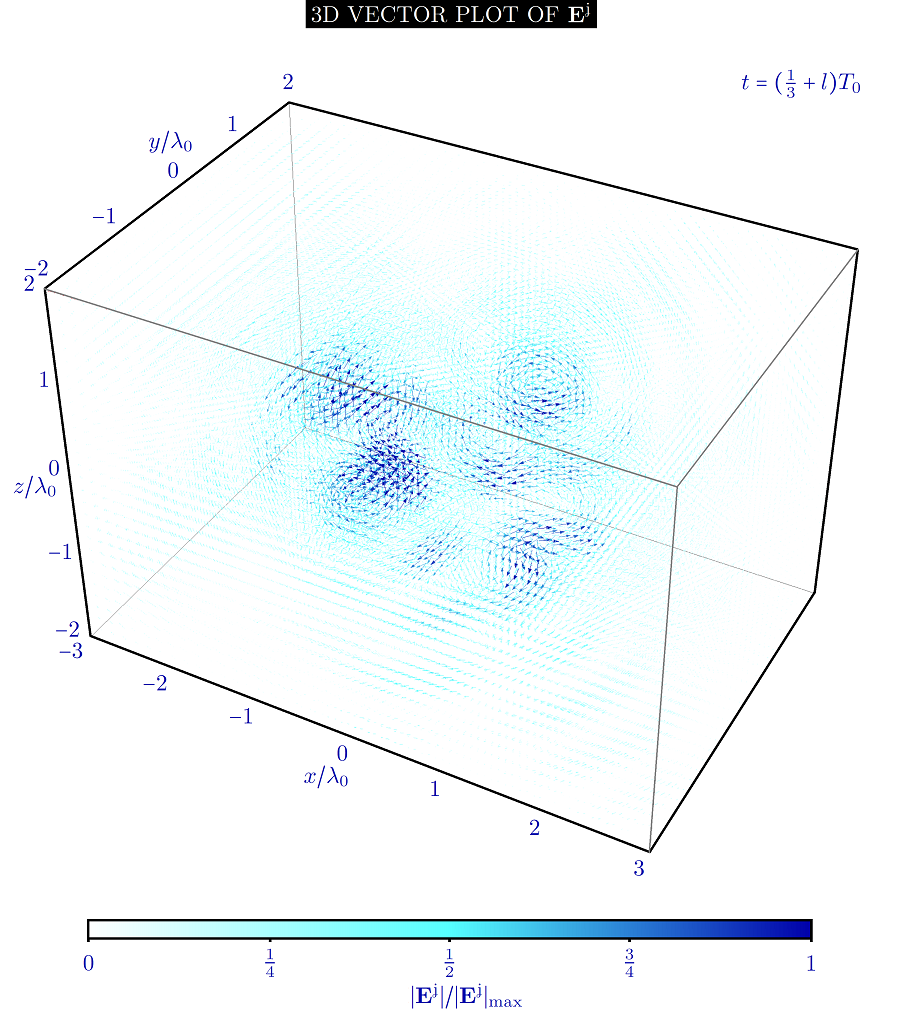}\\
\caption{\small The electric field $\mathbf{E}^\textrm{\textipa{\textctj}}$ of our electromagnetic cloud (section \ref{Clouds}) for $t=(1/3+l)T_0$, depicted as a three-dimensional vector plot: each blue arrow represents an electric field vector, colour-coded and scaled by magnitude.} 
\label{Cloud3Dt1}
\end{figure}

\begin{figure}[h!]
\centering
\includegraphics[width=\linewidth]{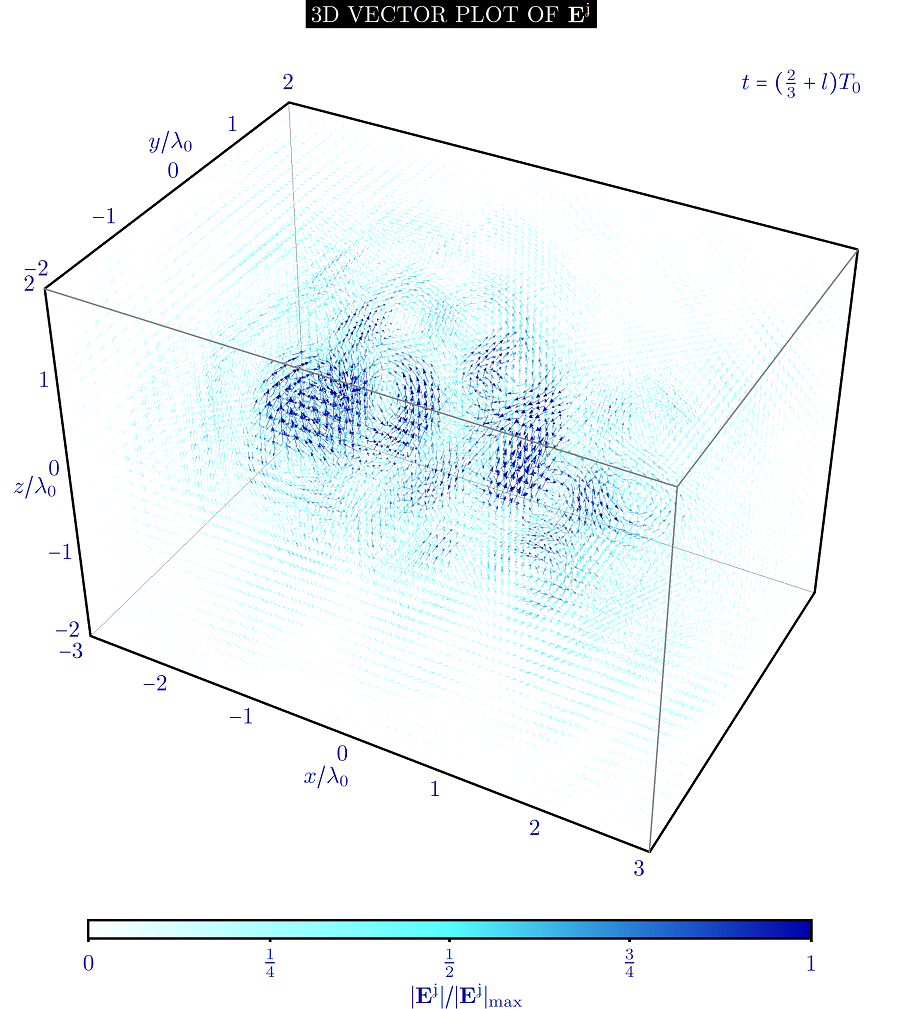}\\
\caption{\small The electric field $\mathbf{E}^\textrm{\textipa{\textctj}}$ of our electromagnetic cloud (section \ref{Clouds}) for $t=(2/3+l)T_0$, depicted as a three-dimensional vector plot: each blue arrow represents an electric field vector, colour-coded and scaled by magnitude.} 
\label{Cloud3Dt2}
\end{figure}

Consider for example the `electromagnetic cloud' specified by taking $N=20$ and
\begin{center}
\begin{tabular}{r r r r r r r r r}
$n$ & $\phi_n$ & $\theta_n$ & $\chi_n$ & $X_n/\lambda_0$ & $Y_n/\lambda_0$ & $Z_n/\lambda_0$  & $\tilde{\mathtt{A}}_{0n}'$  & $\tilde{\mathtt{B}}_{0n}'$ \\  
$1$ & $0.20$ & $1.60$ & $5.60$ & $0.00$ & $0.00$ & $0.69$  & $0.10\textrm{e}^{3.10\textrm{i}}$  & $0.90\textrm{e}^{4.70\textrm{i}}$  \\  
$2$ & $1.40$ & $0.70$ & $6.10$ & $-0.64 $ & $0.15$ & $0.15$  & $0.60\textrm{e}^{1.10\textrm{i}}$  & $0.20\textrm{e}^{1.10\textrm{i}}$  \\  
$3$ & $4.90$ & $0.00$ & $0.30$ & $-1.63$ & $0.57$ & $-0.30$  & $1.20\textrm{e}^{0.60\textrm{i}}$  & $0.70\textrm{e}^{0.60\textrm{i}}$  \\  
$4$ & $3.00$ & $1.20$ & $2.40$ & $-0.91$ & $0.21$ & $-0.15$  & $0.10\textrm{e}^{4.10\textrm{i}}$  & $0.50\textrm{e}^{5.10\textrm{i}}$  \\  
$5$ & $5.20$ & $3.10$ & $5.10$ & $-0.82$ & $-0.18$ & $-0.48$  & $0.70\textrm{e}^{2.90\textrm{i}}$  & $1.20\textrm{e}^{4.10\textrm{i}}$  \\  
$6$ & $0.80$ & $2.80$ & $3.80$ & $-1.30$ & $-0.09$ & $0.36$  & $0.50\textrm{e}^{0.80\textrm{i}}$  & $1.00\textrm{e}^{3.10\textrm{i}}$  \\  
$7$ & $2.20$ & $2.30$ & $4.20$ & $-0.55$ & $-0.42$ & $0.12$  & $0.90\textrm{e}^{1.20\textrm{i}}$  & $0.80\textrm{e}^{2.50\textrm{i}}$  \\  
$8$ & $3.70$ & $0.80$ & $1.30$ & $-1.27$ & $-0.60$ & $0.39$  & $1.10\textrm{e}^{0.10\textrm{i}}$  & $1.10\textrm{e}^{6.00\textrm{i}}$  \\  
$9$ & $2.40$ & $1.40$ & $4.40$ & $-0.94$ & $0.30$ & $-0.33$  & $0.20\textrm{e}^{5.40\textrm{i}}$  & $0.40\textrm{e}^{0.90\textrm{i}}$  \\  
$10$ & $4.30$ & $1.90$ & $0.90$ & $-0.46$ & $0.45$ & $0.06$  & $0.30\textrm{e}^{0.20\textrm{i}}$  & $0.20\textrm{e}^{1.10\textrm{i}}$  \\  
$11$ & $1.20$ & $2.80$ & $4.50$ & $0.70$ & $0.15$ & $0.66$  & $0.80\textrm{e}^{4.10\textrm{i}}$  & $0.40\textrm{e}^{3.20\textrm{i}}$  \\  
$12$ & $0.30$ & $1.60$ & $1.10$ & $0.64$ & $-0.18$ & $0.24$  & $0.50\textrm{e}^{2.50\textrm{i}}$  & $1.00\textrm{e}^{2.30\textrm{i}}$  \\  
$13$ & $1.80$ & $1.30$ & $5.30$ & $1.24$ & $0.48$ & $0.30$  & $0.60\textrm{e}^{1.70\textrm{i}}$  & $0.40\textrm{e}^{2.60\textrm{i}}$  \\  
$14$ & $4.50$ & $0.90$ & $6.10$ & $1.15$ & $0.12$ & $-0.18$  & $1.10\textrm{e}^{3.50\textrm{i}}$  & $0.90\textrm{e}^{6.10\textrm{i}}$  \\  
$15$ & $6.00$ & $0.10$ & $2.30$ & $1.03$ & $-0.63$ & $-0.39$  & $0.90\textrm{e}^{2.40\textrm{i}}$  & $0.10\textrm{e}^{3.10\textrm{i}}$  \\  
$16$ & $0.40$ & $1.80$ & $2.20$ & $0.94$ & $-0.24$ & $0.51$  & $0.60\textrm{e}^{0.20\textrm{i}}$  & $0.20\textrm{e}^{6.20\textrm{i}}$  \\  
$17$ & $1.40$ & $3.10$ & $3.20$ & $1.48$ & $0.39$ & $-0.21$  & $0.40\textrm{e}^{0.50\textrm{i}}$  & $1.10\textrm{e}^{5.10\textrm{i}}$  \\  
$18$ & $3.20$ & $2.80$ & $1.80$ & $1.12$ & $0.51$ & $-0.09$  & $0.10\textrm{e}^{5.50\textrm{i}}$  & $0.60\textrm{e}^{1.70\textrm{i}}$  \\  
$19$ & $1.90$ & $1.50$ & $3.60$ & $0.52$ & $-0.30$ & $0.45$  & $0.20\textrm{e}^{5.30\textrm{i}}$  & $0.50\textrm{e}^{0.50\textrm{i}}$  \\  
$20$ & $5.10$ & $2.40$ & $0.90$ & $0.64$ & $0.21$ & $0.66$  & $0.80\textrm{e}^{0.50\textrm{i}}$  & $1.10\textrm{e}^{1.40\textrm{i}}$  \\  
\end{tabular}
\end{center}
in (\ref{ESECOND}) and (\ref{BSECOND}): our electromagnetic cloud can be regarded as a modest superposition of spatiotemporally translated and rotated versions of our electric ring (\ref{Rings}) and our electric globule (\ref{Globules}), with the translations and rotations assigned randomly (within appropriate ranges). The well localised appearances of the rings and globules see the cloud itself exhibit a well localised appearance. The random choices nevertheless give the cloud an amorphous form, with the phase differences between the rings and globules giving the cloud a more intricate temporal dependence within each cycle than that exhibited by each of the rings and globules individually. The electric field $\mathbf{E}^\textrm{\textipa{\textctj}}$ of our electromagnetic cloud is depicted at different times in Fig. \ref{Cloud3Dt0}, Fig. \ref{Cloud3Dt1} and Fig. \ref{Cloud3Dt2}. The magnetic field $\mathbf{B}^\textrm{\textipa{\textctj}}$ of our electromagnetic cloud is similarly cloud-like.

\begin{figure}[h!]
\centering
\includegraphics[width=\linewidth]{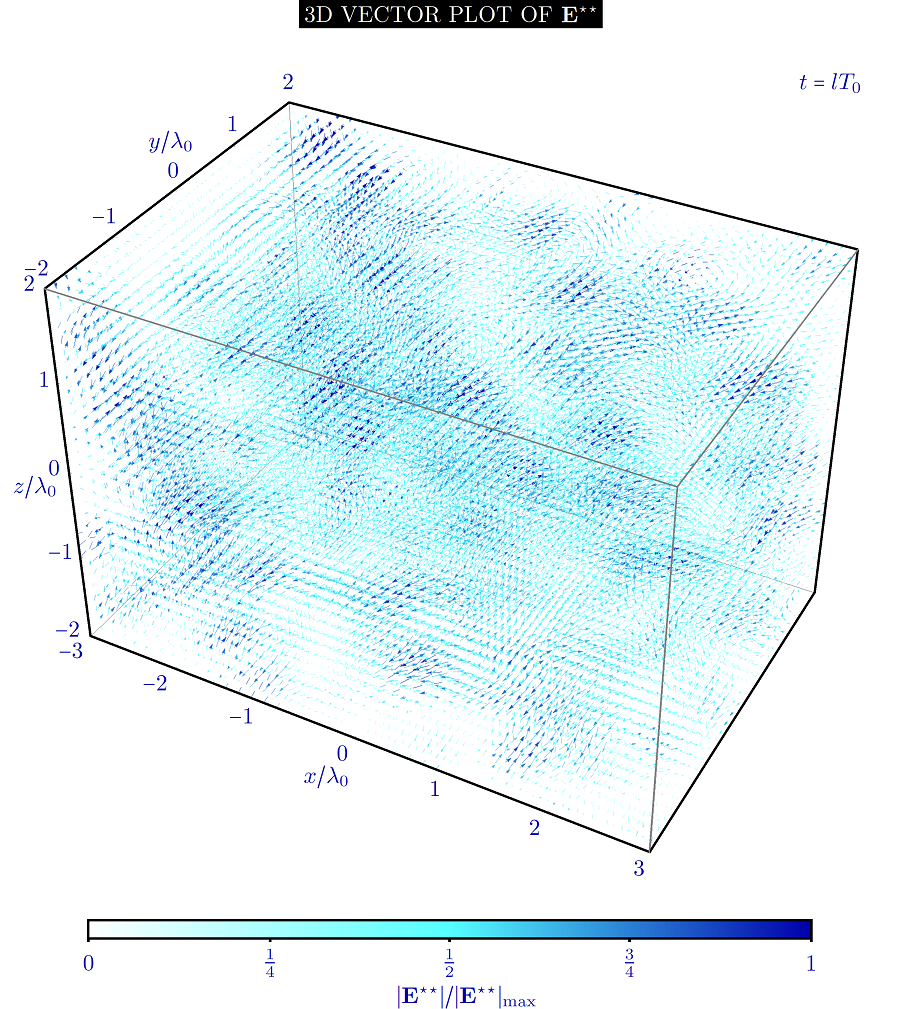}\\
\caption{\small The electric field $\mathbf{E}^{\star\star}$ of a random superposition of plane electromagnetic waves (section \ref{Clouds}), depicted as a three-dimensional vector plot at an instant in time: each blue arrow represents an electric field vector, colour-coded and scaled by magnitude.} 
\label{Waves3D}
\end{figure}

Let us emphasise that our electromagnetic cloud is not merely a random superposition of plane electromagnetic waves (although such superpositions are interesting in their own right \cite{Connor87a}). In general such superpositions do not yield well localised electromagnetic disturbances. Rather, our electromagnetic cloud is a randomly chosen superposition of electric rings and electric globules, each of which is a \textit{carefully} chosen superposition of plane waves. The electric field $\mathbf{E}^{\star\star}$ of a random superposition of plane electromagnetic waves is depicted in Fig. \ref{Waves3D}. Comparing this with Fig. \ref{Cloud3Dt0}, Fig. \ref{Cloud3Dt1} and Fig. \ref{Cloud3Dt2}, it can be seen immediately that there is indeed a basic distinction between our electromagnetic cloud and such superpositions. $\mathbf{E}^{\star\star}$ was generated by taking $\tilde{\mathtt{A}}'=2\pi\sum_{n=1}^N\tilde{\mathtt{A}}_{n0}'\delta(\varphi-\varphi_n) \delta(\vartheta-\vartheta_n)/\sin\vartheta_n$
and $\tilde{\mathtt{B}}'=2\pi\sum_{n=1}^N\tilde{\mathtt{B}}_{n0}'\delta(\varphi-\varphi_n) \delta(\vartheta-\vartheta_n)/\sin\vartheta_n$ in (\ref{Emono}) and (\ref{Bmono}), together with the values of $N$, the $\phi_n$, the $\theta_n$, the $\tilde{\mathtt{A}}_{n0}'$ and the $\tilde{\mathtt{B}}_{n0}'$ listed above for our electromagnetic cloud (which we recycle here for the sake of brevity).

\newpage

\newpage

\newpage

\newpage


\newpage
\section{Energy and temporal dependence}
\label{Energyetc}
In the present section we make some general observations with regards to the energies and temporal dependencies of our unusual electromagnetic disturbances. 

Some of our unusual electromagnetic disturbances (such as our electric ring (section \ref{Rings}), our electric globule (section \ref{Globules}), our electromagnetic tangle (section \ref{Roses}), our electric loop (section \ref{Loops}), our electric link (section \ref{Links}), our electric lines (section \ref{Lines}) and our electric knots (section \ref{REALKnots})) can be regarded as (exotic) electromagnetic standing waves: for both the electric field $\mathbf{E}$ and the magnetic field $\mathbf{B}$, the spatial and temporal dependencies factorise, with the temporal dependence consisting simply of a sinusoidal oscillation. Our other unusual electromagnetic disturbances (such as our electromagnetic cloud (section \ref{Clouds})) cannot be regarded as standing waves, however, due to their more intricate temporal dependencies.

For each of our unusual electromagnetic disturbances that can be regarded as a standing wave, one can say that there is no \textit{flow} of energy on average in any direction at any position, in that the integral of the familiar electromagnetic energy flux density \cite{Griffiths99,Jackson99,Bessel-Hagen21}
\begin{equation}
\frac{1}{\mu_0}\mathbf{E}\times\mathbf{B} \nonumber
\end{equation}
over one period $T_0$ vanishes everywhere for any initial time $t_0$:
\begin{equation}
\frac{1}{T_0}\int_{t_0}^{t_0+T_0}\frac{1}{\mu_0}\mathbf{E}\times\mathbf{B} \textrm{d}t=0.
\end{equation}
One can also say, however, that there is an \textit{oscillation} of energy, back and forth between the electric field, which exhibits a time dependence of the form $\mathbf{E}\propto \cos (\omega_0 t+\delta_0)$, and the magnetic field, which exhibits a time dependence of the form $\mathbf{B}\propto \sin (\omega_0 t+\delta_0)$, thus differing in phase from the electric field by a quarter cycle. When $t= (l -\delta_0/\pi)T_0/2$, $|\mathbf{E}|$ takes its maximum value everywhere whereas $\mathbf{B}=0$, in which case the familiar electromagnetic energy density \cite{Cohen89,Griffiths99,Jackson99,Bessel-Hagen21}
\begin{equation}
\frac{1}{2}\left(\epsilon_0 |\mathbf{E}|^2+\frac{1}{\mu_0}|\mathbf{B}|^2\right) \nonumber
\end{equation}
reduces to
\begin{equation}
\frac{\epsilon_0}{2}|\mathbf{E}|^2, \nonumber
\end{equation}
indicating that energy is stored in the electric field. When $t=(1/2+l -\delta_0/\pi)T_0/2$, $\mathbf{E}=0$ whereas $|\mathbf{B}|$ takes its maximum value everywhere, in which case the familiar electromagnetic energy density reduces to
\begin{equation}
\frac{1}{2\mu_0}|\mathbf{B}|^2, \nonumber
\end{equation}
indicating that energy is stored in the magnetic field instead. Such oscillations are, perhaps, a hallmark of electromagnetic standing waves and are sustained throughout the disturbance by electromagnetic induction. For each of our unusual electromagnetic disturbances that cannot be regarded as a standing wave the situation is different in that
\begin{equation}
\frac{1}{T_0}\int_{t_0}^{t_0+T_0}\frac{1}{\mu_0}\mathbf{E}\times\mathbf{B} \textrm{d}t\ne0
\end{equation}
in general and there is thus a sense in which the disturbance is propagating, at least locally. 

The above allows leads us to draw a simple distinction between each of our unusual electromagnetic disturbances that can be regarded as a standing wave and the region of heightened `intensity' found in a tightly focussed laser beam, for example: one can say that the former do not transport energy on average whereas a beam of light does transport energy on average, in particular through planes perpendicular to its direction of propagation. 

As highlighted in section \ref{Introduction}, each of our unusual electromagnetic disturbances appears to be well localised in all three spatial dimensions in that the electric (and magnetic) field strengths in the regions described are significantly larger than those found anywhere else. It should be noted, however, that the total energy of each of our unusual electromagnetic disturbances (as we have defined them thus far in the present paper) is infinite in that the integral of the familiar electromagnetic energy density over all space diverges:
\begin{equation}
\int_{-\infty}^{\infty}\int_{-\infty}^{\infty}\int_{-\infty}^{\infty}\frac{1}{2}\left(\epsilon_0 |\mathbf{E}|^2+\frac{1}{\mu_0}|\mathbf{B}|^2\right)\textrm{d}x\textrm{d}y\textrm{d}z=\infty. \label{explosion}
\end{equation}
Similar observations can be made of other, well-known solutions of Maxwell's equations (\ref{DivE})-(\ref{Maxwell}): a monochromatic Bessel beam has an infinite energy per unit length \cite{Durnin 87a}; a monochromatic Gaussian beam has a finite energy per unit length but also an infinite total energy, as seen when integrating the energy per unit length along the direction of propagation. Good approximations to these solutions are nevertheless generated routinely within finite regions of space in the laboratory \cite{Durnin 87b}. A subtlety is that these approximations are only quasi-monochromatic, as they exist over finite time intervals. This leads us to recognise that the total energies of our unusual electromagnetic disturbances can be rendered meaningful by introducing suitable regularisations into the frequency spectra of the disturbances. For each of our unusual electromagnetic disturbances of the first kind (section \ref{The first kind}), one such regularisation consists of taking $\tilde{\mathtt{A}}=\textrm{rect}[(k-k_0)/2\delta k]\tilde{\mathtt{A}}'_0/4\pi k^2\delta k$ and $\tilde{\mathtt{B}}=\textrm{rect}[(k-k_0)/2 \delta k]\tilde{\mathtt{B}}'_0/4\pi k^2\delta k$ rather than $\tilde{\mathtt{A}}=E_0\delta(k-k_0)\tilde{\mathtt{A}}'_0/2\pi k_0^2$ and $\tilde{\mathtt{B}}=E_0\delta(k-k_0)\tilde{\mathtt{B}}'_0/2\pi k_0^2$ in (\ref{Emono}) and (\ref{Bmono}), where $\textrm{rect}(K)$ is the rectangular function (with $\textrm{rect}^2(K)=\textrm{rect}(K)$) and $0<2\delta k\ll k_0$ is a range of wavenumbers. This renders the total energy finite and time-independent at finite times ($|t|<\infty$), as desired:
\begin{eqnarray}
&&\int_{-\infty}^{\infty}\int_{-\infty}^{\infty}\int_{-\infty}^{\infty}\frac{1}{2}\left(\epsilon_0 |\mathbf{E}|^2+\frac{1}{\mu_0}|\mathbf{B}|^2\right)\textrm{d}x\textrm{d}y\textrm{d}z \label{totalenergyfinite} \\
&=&\int_0^{2\pi}\int_0^\pi\int_0^\infty4\pi^3\epsilon_0\left(|\tilde{\mathtt{A}}|^2+|\tilde{\mathtt{B}}|^2\right) k^2 \textrm{d}k\sin\vartheta\textrm{d}\vartheta\textrm{d}\varphi \nonumber \\
&=&\frac{2\pi^2\epsilon_0}{\left(k_0^2-\delta k^2\right)\delta k}\left(|\tilde{\mathtt{A}}_0'|^2+|\tilde{\mathtt{B}}_0'|^2\right). \nonumber 
\end{eqnarray}
One can appreciate this regularisation as follows. Recall from section \ref{fgh} that the modulating scalar fields $f$, $g$ and $h$ each tend towards a form that is at least qualitatively similar to a sinusoidal undulation modulated by a $1/|\mathbf{r}|$ fall off as $|\mathbf{r}|\rightarrow\infty$ in any non-trivial direction. Let us therefore consider
\begin{equation}
\frac{1}{|\mathbf{r}|}\cos\left(k_0|\mathbf{r}|\right)\cos\left(c k_0 t\right) \label{style}
\end{equation}
for the sake of concreteness. Our regularisation corresponds to replacing this with
\begin{equation}
\frac{1}{2\delta k}\int_{k_0-\delta k}^{k_0+\delta k}\frac{1}{|\mathbf{r}|}\cos\left(k' |\mathbf{r}|\right)\cos\left(c k' t\right) \textrm{d}k', \label{stylez} 
\end{equation}
which for finite times tends towards
\begin{eqnarray}
&&\frac{1}{\delta k_0|\mathbf{r}|^2}[\cos(k_0|\mathbf{r}|)\cos(ck_0t)\sin(\delta k|\mathbf{r}|)\cos(\delta k c t) \label{stylezz} \\
&&-\sin(k_0|\mathbf{r}|)\sin(ck_0t)\sin(\delta k ct)\cos(\delta k|\mathbf{r}|)]   \nonumber
\end{eqnarray}
as $|\mathbf{r}|\rightarrow \infty$. The $1/|\mathbf{r}|^2$ fall off seen in (\ref{stylezz}) is more dramatic than the $1/|\mathbf{r}|$ fall off seen in (\ref{style}), as desired. More potent regularisations might be required to render quantities like the total angular momentum \cite{Cohen89,Griffiths99,Jackson99,Bessel-Hagen21}
\begin{equation}
\int_{-\infty}^{\infty}\int_{-\infty}^{\infty}\int_{-\infty}^{\infty}\epsilon_0\mathbf{r}\times(\mathbf{E}\times\mathbf{B})\textrm{d}^3\mathbf{r} \nonumber
\end{equation}
and the total boost angular momentum \cite{Bessel-Hagen21}
\begin{eqnarray}
&&t\int_{-\infty}^{\infty}\int_{-\infty}^{\infty}\int_{-\infty}^{\infty}\epsilon_0(\mathbf{E}\times\mathbf{B})\textrm{d}^3\mathbf{r} \\
&&-\frac{1}{c^2}\int_{-\infty}^{\infty}\int_{-\infty}^{\infty}\int_{-\infty}^{\infty}\frac{1}{2}\mathbf{r}\left(\epsilon_0|\mathbf{E}|^2+\frac{1}{\mu_0}|\mathbf{B}|^2\right)\textrm{d}^3\mathbf{r} \nonumber
\end{eqnarray}
meaningful, due to the presence of $\mathbf{r}$ in the integrands.

The author acknowledges that densities and flux densities are not necessarily unique \cite{Cameron15}. The `familiar' forms considered above enjoy something of a privileged status, however, in that they seem to be singled out by gravitational interactions.

Let us conclude here by noting with interest that each of our unusual electromagnetic disturbances can be set translating with any speed $<c$ in any direction by actively boosting it \cite{Griffiths99,Jackson99,Einstein05}. We are free to do this because (\ref{DivE})-(\ref{Maxwell}) are Lorentz-invariant \cite{Bessel-Hagen21}. For speeds $\ll c$, this does not significantly alter the basic form of the disturbance. It is thus possible to have an ultraviolet \cite{UV} electromagnetic tangle translating at $10$m.s$^{-1}$, for example. Each of the plane electromagnetic waves comprising a boosted version of one of our unusual electromagnetic disturbances propagates with speed $c$ \cite{Cohen89,Griffiths99,Jackson99,Maxwell65,Einstein05,Giovannini15}. It is the differences in the frequencies of these waves that gives rise to the apparent translation of the disturbance as a whole: the boosted version of a monochromatic electromagnetic disturbance is polychromatic in general, its constituent plane waves having been Doppler shifted \cite{Jackson99,Doppler42} by different amounts depending upon their direction of propagation with respect to the direction of the boost.


\newpage
\section{Generation}
\label{Generation}
In the present section we consider the generation of our unusual electromagnetic disturbances in the laboratory. Different methods of generation will be required depending upon the form of disturbance sought and the frequencies required. In section \ref{VinDiesel} we focus upon the use of an antenna to generate an electric ring (section \ref{Rings}) in the radiowave or microwave domain \cite{Maxwell65}. In section \ref{VinDiesel2} we focus upon the use of cylindrically polarised vector beams to generate an electric ring or electric globule (section \ref{Globules}) in the visible domain.

Works to date on electromagnetic knots are largely silent on the question of generation \cite{Trautman77, Ranada89, Ranada90,Ranada92, Ranada95, Ranada97,  Ivo04, Besieris09, Arrayas10, Irvine10, Arrayas11, Ranada12, Arrayas12, Dalhuisen12, vanEnk13, Kedia13, Swearngin14, Arrayas15, Thompson15, Hoyos15, Kholodenko16b, Kholodenko16a, Kedia16, Arrayas17b, Smith17, Alves17, Chubykalo02,Arnhoff92}, although some basic discussions have been presented \cite{Irvine08,Arrayas17a}. 

\subsection{Radiowave or microwave domain} 
\label{VinDiesel}
\begin{figure}[h!]
\centering
\includegraphics[width=\linewidth]{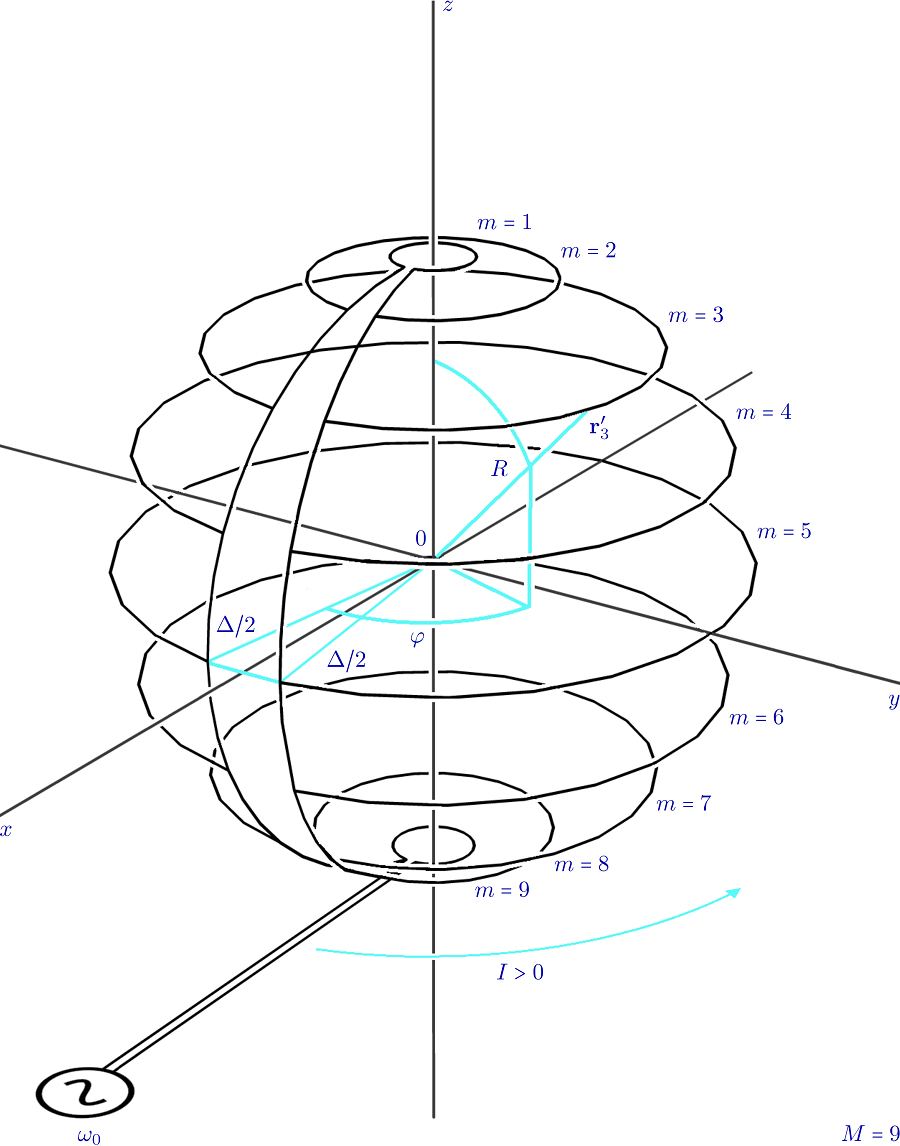}\\
\caption{\small The basic ingredients of our electric-ring antenna (section \ref{VinDiesel}).} 
\label{Generator1}
\end{figure}

Consider a collection of $M$ electrically conducting rings, with each ring concentric with the surface of a sphere of radius $R$ centred upon the origin; that section of each ring with $\varphi\in(-\Delta/2,\Delta/2)$ removed; the ends of each ring connected to an alternating electric current source of (central) angular frequency $\omega_0$; $\theta_m$ the polar angle of the $m$th ring ($m\in\{1,\dots,M\}$) and
\begin{equation}
I=\Re\left(\tilde{I}\textrm{e}^{-\textrm{i}\omega_0 t}\right)
\end{equation}
the (identical) current in each ring, directed in the $\hat{\pmb{\phi}}$ direction for $I>0$. The basic ingredients of our electric-ring antenna are depicted in Fig. \ref{Generator1}. Ignoring interactions between rings, taking each ring to be of negligible cross-section, neglecting the radiation produced by the elements that connect the rings to the current source and assuming the surrounding medium to be transparent with phase refractive index $n_p$, we find that the electric field radiated by the rings is essentially
\begin{equation}
\mathbf{E}^\textrm{LF}=\Re\left(\tilde{\mathbf{E}}^\textrm{LF}\textrm{e}^{-\textrm{i}\omega_0 t}\right)
\end{equation}
with
\begin{eqnarray}
&&\tilde{E}_a^\textrm{LF}= \frac{\textrm{i} \mu_0 R \omega_0 \tilde{I} }{4 \pi} \sum_{m=1}^M \int_{\Delta/2}^{2 \pi -\Delta /2} \frac{\textrm{e}^{\textrm{i}k_0 n_p |\mathbf{r}-\mathbf{r}_m'|}}{|\mathbf{r}-\mathbf{r}_m'|}\sum_{b\in\{x,y,z\}}   \\
&& \Bigg [\delta_{ab}
\Bigg(1+\frac{\textrm{i}}{k_0 n_p |\mathbf{r}-\mathbf{r}_m'|}-\frac{1}{k_0^2 n_p^2 |\mathbf{r}-\mathbf{r}_m'|^2}\Bigg)-\nonumber \\
&&\frac{\left(\mathbf{r}-\mathbf{r}_m'\right)_a\left(\mathbf{r}-\mathbf{r}_m'\right)_b}{|\mathbf{r}-\mathbf{r}_m'|^2}\Bigg(1+\frac{\textrm{3 i}}{k_0 n_p |\mathbf{r}-\mathbf{r}_m'|}-\frac{3}{k_0^2 n_p^2 |\mathbf{r}-\mathbf{r}_m'|^2}\Bigg) \Bigg]\hat{\varphi}_b \sin\vartheta_m \textrm{d}\varphi, \nonumber
\end{eqnarray}
where $a\in\{x,y,z\}$, $\delta_{ab}$ is the Kronecker delta function and
\begin{equation}
\mathbf{r}_m'=R\left(\hat{\pmb{x}}\cos\varphi\sin\vartheta_m+\hat{\pmb{y}}\sin\varphi\sin\vartheta_m+\hat{\pmb{z}}\cos\vartheta_m\right)
\end{equation}
is the position of an element of the $m$th ring: $\mathbf{E}^\textrm{LF}$ is a sum over contributions due to such elements, with each element treated as an oscillating electric dipole \cite{Griffiths99, Jackson99}.

\begin{figure}[h!]
\centering
\includegraphics[width=\linewidth]{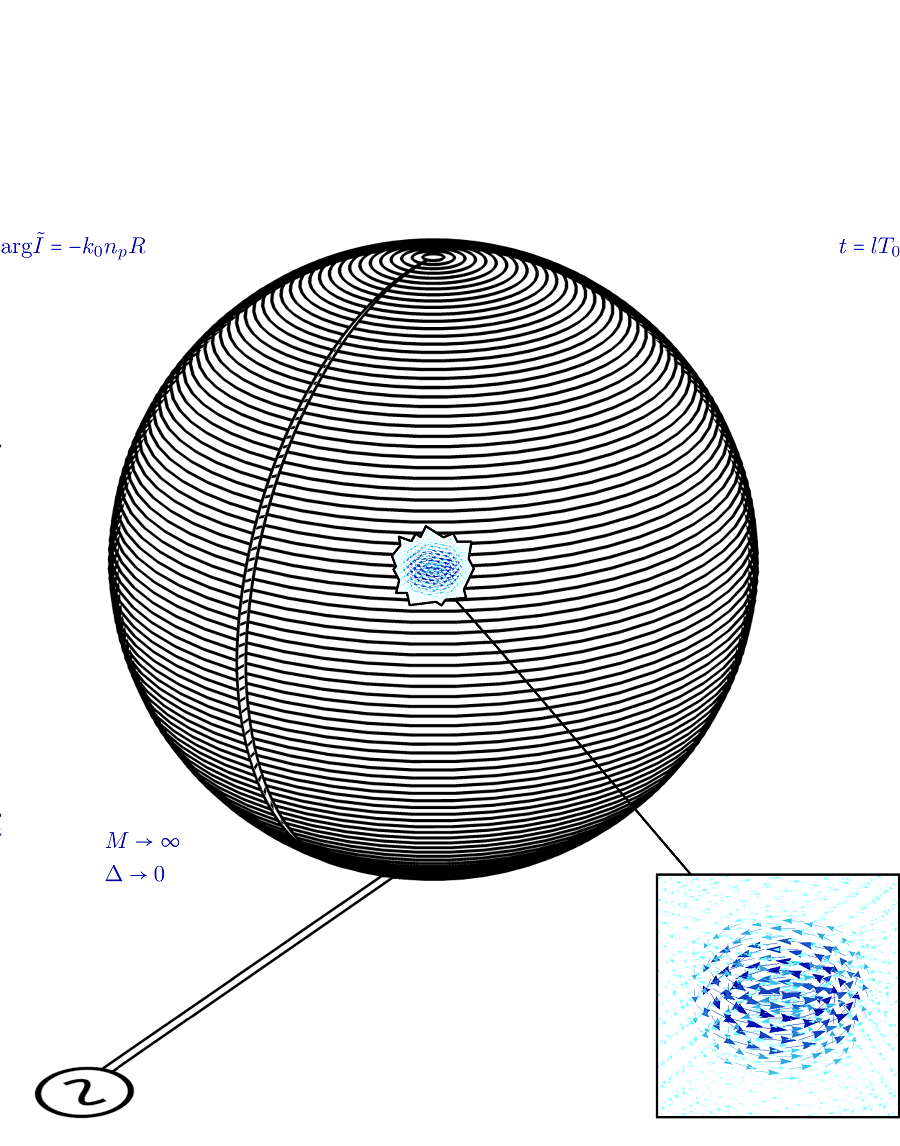}\\
\caption{\small A schematic depiction of our electric-ring antenna (section \ref{VinDiesel}) in operation. Part of the antenna has been cut away to reveal the electric ring (section \ref{Rings}) generated by the antenna.} 
\label{Generator2}
\end{figure}

Suppose now that there is a small gap and many rings such that
\begin{eqnarray}
&&\int_{\Delta /2}^{2\pi-\Delta /2}\textrm{d}\varphi \rightarrow\int_0^{2\pi}\textrm{d}\varphi \nonumber \\
&&\vartheta_m\rightarrow \vartheta, \nonumber \\
&&\sum_{m=1}^M\rightarrow\frac{M}{\pi}\int_0^\pi \textrm{d}\vartheta. \nonumber
\end{eqnarray}
For regions within the sphere ($|\mathbf{r}|< R$) that satisfy the far-field condition ($k_0 n_p \left(R-|\mathbf{r}|\right)\gg 1$),
\begin{eqnarray}
&&\tilde{E}_a^\textrm{LF}\approx \frac{\textrm{i}\mu_0R\omega_0\tilde{I}M}{4 \pi^2}\int_0^{2\pi}    \int_0^{\pi } \frac{\textrm{e}^{\textrm{i}k_0 n_p|\mathbf{r}-\mathbf{r}'|}}{|\mathbf{r}-\mathbf{r}'|}  \\
&& \Bigg [\delta_{ab} -\frac{\left(\mathbf{r}-\mathbf{r}'\right)_a\left(\mathbf{r}-\mathbf{r}'\right)_b}{|\mathbf{r}-\mathbf{r}'|^2}  \Bigg] \hat{\varphi}_b \sin\vartheta\textrm{d}\vartheta\textrm{d}\varphi  \nonumber
\end{eqnarray}
where
\begin{equation}
\mathbf{r}'=R\left(\hat{\pmb{x}}\cos\varphi\sin\vartheta+\hat{\pmb{y}}\sin\varphi\sin\vartheta+\hat{\pmb{z}}\cos\vartheta\right).
\end{equation}
If, moreover, the distance from the origin is small ($|\mathbf{r}|\ll R$) such that
\begin{eqnarray}
&&\frac{\textrm{e}^{\textrm{i}k_0 n_p |\mathbf{r}-\mathbf{r}'|}}{|\mathbf{r}-\mathbf{r}'|} \approx \frac{\textrm{e}^{\textrm{i}k_0 n_p R}\textrm{e}^{-\textrm{i}k_0 n_p \mathbf{r}'\cdot\mathbf{r}/R}}{R} \\
&&\left[\delta_{ab}-\frac{\left(\mathbf{r}-\mathbf{r}'\right)_a\left(\mathbf{r}-\mathbf{r}'\right)_b}{|\mathbf{r}-\mathbf{r}'|^2}\right]\hat{\varphi}_b \approx \hat{\varphi}_a,
\end{eqnarray}
this reduces further to
\begin{eqnarray}
\tilde{\mathbf{E}}^\textrm{LF}\approx  \tilde{E}_0'\hat{\pmb{\phi}} f'    \textrm{e}^{-\textrm{i}\omega_0 t}
\end{eqnarray}
with
\begin{eqnarray}
\tilde{E}_0' &=& \frac{\mu_0 \omega_0 \tilde{I}M\textrm{e}^{\textrm{i} k_0 n_p R}}{2\pi} \\
f' &=& \int_0^\pi \textrm{J}_1(k_0\sin\vartheta n_p s ) \cos (k_0\cos\vartheta n_p z) \sin \vartheta \textrm{d}\vartheta.
\end{eqnarray}
Thus
\begin{equation}
\mathbf{E}^\textrm{LF}\approx|\tilde{E}_0'|\hat{\pmb{\phi}}f'\cos(\omega_0 t-\textrm{arg}\tilde{E}_0'),
\end{equation}
which is essentially the electric field $\mathbf{E}^\textrm{\textipa{\textscripta}}$ of our electric ring: the two coincide precisely in form for $n_p\approx 1$ so that $f'\approx f$, together with a choice of phase for $I$ such that $\tilde{E}_0' \rightarrow E_0$ is real. Our electric-ring antenna is depicted schematically in operation in Fig. \ref{Generator2}. 

The geometrical requirements above are reasonably well satisfied by $M=100$, $R=1.0\times10^{-1}$m and $\omega_0/2\pi=1.3\times10^{10}$s$^{-1}$, for example. 

Simple elaborations upon our design permit the generation of other unusual electromagnetic disturbances in the radiowave or microwave domain.

\newpage

\newpage

\newpage
\subsection{Visible domain} 
\label{VinDiesel2}
The following observations were borne out of discussions with Sonja Franke-Arnold, Neal Radwell and Fiona C Speirits, to whom the author is grateful.

The electric field $\mathbf{E}^\textrm{\textipa{\textscripta}}$ of our electric ring can be recast as
\begin{equation}
\mathbf{E}^\textrm{\textipa{\textscripta}}=\frac{1}{k_0}\int_0^{k_0} \left[\mathbf{E}_+(|k_z|)+\mathbf{E}_-(|k_z|)\right]\textrm{d}|k_z|
\end{equation}
with
\begin{equation}
\mathbf{E}_\pm(|k_z|)=\Re\left[E_0\hat{\pmb{\phi}}\textrm{J}_1(\kappa s)\textrm{e}^{\textrm{i}\left( \pm|k_z|z-\omega_0t\right)}\right]
\end{equation}
an azimuthally polarised vector beam \cite{Radwell16, Ornigotti16,Chille16, Radwell16b,Pohl 72a, Zhan 09a} of transverse wavenumber $\kappa=\sqrt{k_0^2-|k_z|^2}$ propagating in the $\pm z$ direction. This leads us to suggest, tentatively, that a good approximation to our electric ring in the visible domain might be generated in the laboratory by superposing counter-propagating azimuthally polarised vector beams of light, each with an appropriate spread of transverse wave numbers. An analogous arrangement with radially polarised \cite{Dunlop17,Radwell16,Ornigotti16, Chille16,Radwell16b,Zhan 09a} beams might be used to generate a good approximation to our electric globule. Electric rings and electric globules created in this way could be used in turn as building blocks to generate more elaborate unusual electromagnetic disturbances in the visible domain, following the recipes given in section \ref{The second kind}, for example.

The tight focussing of radially polarised vector beams has been explored in considerable detail elsewhere, owing to the fact that such beams can be focussed more tightly than usual \cite{Zhan 09a,Quabis00, Dorn03}. It is interesting to note that our electric globule, which can be regarded as a superposition of radially polarised vector beams as described above, seems to reside within a sphere of sub-wavelength diameter, in accord with the aforementioned focussing properties of the beams individually. The tight focussing of an azimuthally polarised vector beam has recently been explored \cite{Porfirev 16a}, with interesting results. 


\newpage
\section{Outlook}
\label{Discuss}
We recognise many possible directions for future research, some of which are highlighted below.

With regards to basic theory:
\begin{itemize}
\item It is desirable to develop more systematic methods for constructing unusual electromagnetic disturbances. This could lead to applications in the arts, for example, with unusual electromagnetic disturbances joining spiral beams \cite{Abramochkin04} and caustics (www.zintaglio.com/lens.html) as a means by which to `paint with light'. Our unusual electromagnetic disturbances would add a new element here in that they appear well localised in all \textit{three} spatial dimensions, not just two. \\

\item The generation of our unusual electromagnetic disturbances in the laboratory requires further consideration. \\

\item Exotic media might permit new types of unusual electromagnetic disturbance \cite{Bouchard16}.
\end{itemize}

With regards to potential applications:
\begin{itemize}
\item Our unusual electromagnetic disturbances might enable new and / or improved forms of display. In particular, unusual electromagnetic disturbances in the visible domain might act as three-dimensional pixels or voxels for a new form of volumetric 3D display, with each voxel rendered visible in air via Rayleigh scattering \cite{Rayleigh71a,Rayleigh71b,Rayleigh71c}. Such voxels would be superposable and non-destructive, the small size of each voxel would permit high-resolution volumetric images, different colours of voxel and therefore images could be realised, the non-isotropic radiation profiles of the voxels would permit emulation of occlusion and opacity and there would be no need for mechanical parts within the image volume. \\

\item The setups that we envisage for the generation of our unusual electromagnetic disturbances in the visible domain are not entirely unlike that found in a `$4\textrm{Pi}$' microscope \cite{Hell 94a}, suggesting possible applications for our unusual electromagnetic disturbances in imaging. \\

\item The use of electromagnetic forces to manipulate various forms of matter, including atoms \cite{Ashkin70, Radwell13, Lembessis16} and fusion plasmas \cite{Klinger17}, is well established. The possibilities offered in such contexts by our unusual electromagnetic disturbances might be worthy of pursuit. A preliminary examination of the forces exerted upon charged particles by an electromagnetic knot is presented in \cite{Arrayas10}. It is particularly interesting to note that the helical magnetic field lines found at appropriate times within each torus of our electromagnetic tangle (section \ref{Roses}) resemble the magnetic field lines engineered in certain tokamaks to confine hot plasma: perhaps our electromagnetic tangle could assist in the pursuit of efficient energy production via controlled thermonuclear fusion. 
\end{itemize}

With regards to natural occurrence:
\begin{itemize}
\item Suggestions have already been made that novel electromagnetic disturbances \cite{Chubykalo02,Arnhoff92}, in particular magnetic knots \cite{Arrayas17a,Ranada96, Ranada98, Ranada00}, might underpin the phenomenon of ball lightning. Perhaps it is worth revisiting these hypotheses given the new ideas introduced in the present paper. Variants of our electric-ring antenna (\ref{Generation}) might be employed to ionise the air in unusual geometries and thus help explore such hypotheses in the laboratory \cite{Chubykalo02, Ohtsuki91, Alexeff95}, for example. \\

\item An unusual electromagnetic disturbance in vacuum might appear essentially invisible when viewed from a sufficient distance, at least as far as its electromagnetic profile is concerned. The disturbance should nevertheless be capable of making its presence known gravitationally, due to its energy-momentum content \cite{Einstein16}. It is interesting to imagine an electromagnetic cloud in interstellar space acting as a gravitational lens, for example. Could some component of dark matter \cite{Cornelius22} be attributed to unusual electromagnetic disturbances and / or analogous features in the gravitational field? The author has found it possible to construct gravitational analogues \cite{Weisberg81, Barnett14, Abbott16} of the unusual electromagnetic disturbances introduced in the present paper, including gravitational clouds.
\end{itemize}

We will return to these and related ideas elsewhere.


\section{Acknowledgments}
The present work was supported by the Engineering and Physical Sciences Research Council (EPSRC) (EP/M004694/1). The author thanks Stephen M Barnett, J\"{o}rg B G\"{o}tte, Sonja Franke-Arnold, Neal Radwell, Gergely Ferenczi, Alison M Yao, Aidan S Arnold, Sophie Viaene, Marco Ornigotti, Megan R Paterson and an anonymous referee for their advice and encouragement.


\newpage
\begin{appendix}

\section{Multipolar electromagnetic waves}
\label{Multipolar}
The following treatment echoes that given in \cite{Cohen89}.

Any electromagnetic disturbance can be regarded as a superposition of multipolar electromagnetic waves: the general solution to Maxwell's equations (\ref{DivE})-(\ref{Maxwell}) can be cast as
\begin{eqnarray}
&&\mathbf{E}=\Re\Bigg\{\int_0^\infty\sum_{J=1}^{\infty}\sum_{M_J=-J}^{J}\textrm{i}\sqrt{\frac{2\hbar c k}{\epsilon_0}}\Big[\tilde{\alpha}_{kJM_JX}\tilde{\mathbf{I}}_{kJM_JX} \\
&&+\tilde{\alpha}_{kJM_JZ}\tilde{\mathbf{I}}_{kJM_JZ}\Big]\textrm{e}^{-\textrm{i}\omega t}\textrm{d}k\Bigg\} \nonumber \\
&&\mathbf{B}=\Re\Bigg\{\int_0^\infty\sum_{J=1}^{\infty}\sum_{M_J=-J}^{J}\frac{\textrm{i}}{c}\sqrt{\frac{2\hbar c k}{\epsilon_0}}\Big[-\tilde{\alpha}_{kJM_JX}\tilde{\mathbf{I}}_{kJM_JZ} \\
&&+\tilde{\alpha}_{kJM_JZ}\tilde{\mathbf{I}}_{kJM_JX}\Big]\textrm{e}^{-\textrm{i}\omega t}\textrm{d}k\Bigg\} \nonumber
\end{eqnarray}
with
\begin{eqnarray}
&&\tilde{\mathbf{I}}_{kJM_JX}(\phi,\theta,r)=\sqrt{\frac{2}{\pi}}k\textrm{i}^J\tilde{\mathbf{X}}_{JM_J}(\phi,\theta)\textrm{j}_J(kr) \\
&&\tilde{\mathbf{I}}_{kJM_JZ}(\phi,\theta,r)=\sqrt{\frac{2}{\pi}}\frac{\textrm{i}^{J-1}}{r}\Bigg\{\sqrt{J(J+1)}\tilde{\mathbf{N}}_{JM_J}(\phi,\theta)\textrm{j}_J(kr) \\
&&+\tilde{\mathbf{Z}}_{JM_J}(\phi,\theta)\frac{\textrm{d}\left[kr\textrm{j}_J(kr)\right]}{\textrm{d}(kr)}\Bigg\}, \nonumber
\end{eqnarray}
as well as
\begin{eqnarray}
\tilde{\mathbf{X}}_{JM_J}(\phi,\theta)&=&\frac{\mathbf{r}\times\pmb{\nabla}\tilde{Y}_{JM_J}(\phi,\theta)}{\sqrt{J(J+1)}}, \\
\tilde{\mathbf{N}}_{JM_J}(\phi,\theta)&=&\hat{\mathbf{r}}\tilde{Y}_{JM_J}(\phi,\theta) \\
\tilde{\mathbf{Z}}_{JM_J}(\phi,\theta)&=&-\hat{\mathbf{r}}\times\tilde{\mathbf{X}}_{JM_J}(\phi,\theta),
\end{eqnarray}
where $\tilde{\alpha}_{kJM_JX}$ and $\tilde{\alpha}_{kJM_JZ}$ are complex functions of $k$, $J\in\{1,\dots\}$ and $M_J\in\{0,\dots,\pm J\}$; $\phi$, $\theta$ and $r$ are spherical coordinates;
\begin{eqnarray}
\hat{\pmb{\phi}}&=&-\hat{\pmb{x}}\sin\phi+\hat{\pmb{y}}\cos\phi, \\
\hat{\pmb{\theta}}&=&\hat{\pmb{x}}\cos\phi\cos\theta+\hat{\pmb{y}}\sin\phi\cos\theta-\hat{\pmb{z}}\sin\theta \\
\hat{\mathbf{r}}&=&\hat{\pmb{x}}\cos\phi\sin\theta+\hat{\pmb{y}}\sin\phi\sin\theta+\hat{\pmb{z}}\cos\theta
\end{eqnarray}
are associated unit vectors; $\textrm{j}_J(X)$ is the spherical Bessel function of order $J$ and $\tilde{Y}_{JM_J}(X,X')$ is the spherical harmonic function of order $J$, $M_J$. A particular disturbance is determined by specifying $\tilde{\alpha}_{kJM_JX}$ and $\tilde{\alpha}_{kJM_JZ}$,  which is equivalent to specifying the normal variables: `$\tilde{\pmb{\alpha}}(\varphi,\vartheta,k,t)=\int_0^\infty\sum_{J=0}^\infty\sum_{M_J=-J}^J\delta(k'-k)(\tilde{\alpha}_{k'JM_JX}\tilde{\mathbf{X}}_{JM_J}(\varphi,\vartheta)+\tilde{\alpha}_{k'JM_JZ}\tilde{\mathbf{Z}}_{JM_J}(\varphi,\vartheta))\textrm{e}^{-\textrm{i}\omega t}\textrm{d}k'/k' $'.

The relationship between the multipolar description outlined above and the plane-wave description used in the main text is embodied by the following relationships:
\begin{eqnarray}
\tilde{\alpha}_{kJM_JX}=-2\textrm{i}\sqrt{\frac{\pi^3\epsilon_0 }{\hbar c}}\int_0^{2\pi}\int_0^\pi \tilde{\mathbf{X}}^\ast_{JM_J}(\varphi,\vartheta)\cdot (\hat{\pmb{\varphi}}\tilde{\mathtt{B}}+\hat{\pmb{\vartheta}}\mathtt{A})\sqrt{k}   \sin\vartheta \textrm{d}\vartheta\textrm{d}\varphi \\
\tilde{\alpha}_{kJM_JZ}=-2\textrm{i}\sqrt{\frac{\pi^3\epsilon_0 }{\hbar c}} \int_0^{2\pi}\int_0^\pi   \tilde{\mathbf{Z}}^\ast_{JM_J}(\varphi,\vartheta)\cdot (\hat{\pmb{\varphi}}\tilde{\mathtt{B}}+\hat{\pmb{\vartheta}}\mathtt{A})\sqrt{k} \sin\vartheta\textrm{d}\vartheta\textrm{d}\varphi.
\end{eqnarray} 
For our unusual electromagnetic disturbances of the first kind (section \ref{The first kind}), we find that $\tilde{\alpha}_{k JM_JX}\propto \delta(k-k_0)\delta_{M_J 0}$ and $\tilde{\alpha}_{kJM_JZ}\propto \delta(k-k_0)\delta_{M_J0}$ with no obvious closed form for the $J$ dependencies.

\end{appendix}



\end{document}